# Alleviating Linguistic and Interactional Anxiety of Non-Native Speakers in Multilingual Communication

PEINUAN QIN, School of Computing, National University of Singapore, Singapore
JUSTIN PENG, School of Computing, National University of Singapore, Singapore
ZHENGTAO XU, School of Computing, National University of Singapore, Singapore
JITING CHENG, School of Computing, National University of Singapore, Singapore
ZICHENG ZHU, School of Computing, National University of Singapore, Singapore
NAOMI YAMASHITA, Social Informatics, Kyoto University, Japan
YI-CHIEH LEE, National University of Singapore, Singapore

Non-native speakers (NNSs) frequently encounter speaking difficulties in multilingual communication, where existing approaches have shown promise in facilitating NNSs' comprehension and participation in real-time communication. However, they often overlook providing direct speaking support, where anxiety stemming from linguistic inadequacy and uncertain communication dynamics are core issues. To address this, we introduce an AI tool with translation for real-time speaking support. It also builds a channel for mutual understanding with native speakers (NSs) to mitigate interactional anxiety. Through a within-subjects experiment involving 25 NNS-NS pairs (N = 50) on collaborative tasks, our findings suggest that the tool improved NNSs' speaking self-efficacy, reduced their interactional anxiety, and decreased their workload, particularly for NNSs with below-average language proficiency. Furthermore, NNSs reported a significant sense of support from their NS partners via the mutual understanding channel, and NSs also clearly perceived the NNSs' need for assistance and displayed a strong sense of communicative responsibility. This research underscores the potential of AI support in real-time NNS communication and the importance of promoting mutual understanding, culminating in actionable design insights for future work.

## 1 Introduction

In global collaboration, it is prevalent to use common languages, such as English, for team communication [39, 95], where non-native speakers (NNSs) often face substantial challenges, particularly in speaking [2, 13, 107]. First, linguistic inadequacies [22, 40, 109] in speaking cause them to fear negative evaluation from their collaborators [48, 55, 76], i.e., linguistic anxiety. Moreover, real-time communication exerts additional pressure on NNSs. For instance, conversational silences or breakdowns [56, 67, 84, 112] can increase NNSs' concerns about the status of conversation, leading to interactional anxiety [37, 68, 104]. These deleterious affective states can inhibit the speaking performance of NNSs and impede their effective contribution to collaborative teamwork [41, 105].

Currently, there is insufficient research [94] focusing directly on supporting NNSs' speaking in real-time communication, resulting in limited empirical evidence for alleviating these two types of anxiety. Existing research mainly aids NNSs through comprehension support or increasing NNS' participation [69, 90]. For instance, Pan et al. [90] allowed native speakers (NSs) to highlight keywords within transcribed speech to reduce NNSs' cognitive load when following complex

Authors' Contact Information: Peinuan Qin, School of Computing, National University of Singapore, Singapore, Singapore, e1322754@u.nus.edu; Justin Peng, School of Computing, National University of Singapore, Singapore, Singapore, justin.peng@u.nus.edu; Zhengtao Xu, School of Computing, National University of Singapore, Singapore, Singapore; Jiting Cheng, School of Computing, National University of Singapore, Singapore, Singapore, e1237249@u.nus.edu; Zicheng Zhu, School of Computing, National University of Singapore, Singapore, Singapore, zicheng@u.nus.edu; Naomi Yamashita, Social Informatics, Kyoto University, Kyoto, Japan; Yi-Chieh Lee, National University of Singapore, Singapore, Singapore, yclee@nus.edu.sg.

conversations. Li et al. [69] developed an agent to monitor turn-taking in communication and intervene when NSs dominated the discussion, thus creating more speaking opportunities for NNSs. In contrast, Qin et al. [94] recently proposed an AI-based speaking assistant that directly empowers NNSs to formulate and verbalize their perspectives in the target language, providing insights for real-time speaking support. However, it failed to reduce interactional anxiety and workload while NNSs seek speaking assistance, resulting in a negative communication experience. This underscores the complexity of supporting NNSs in real-time speaking, suggesting the need to not only alleviate linguistic barriers but also reduce interactional anxiety and workload.

To this aim, we developed a tool comprising two components. First, it aims to address NNSs' linguistic barriers through a speaking assistant, which provides a translation feature [86, 100], allowing NNSs to express themselves in their first language (L1) when they experience difficulties articulating thoughts in their second language (L2). For flexibility and responsiveness, this feature accepts a combination of voice and typed inputs [97]. Moreover, it aims to reduce NNSs' interactional anxiety through a mutual understanding channel. Previous studies claimed that vulnerability can evoke prosocial actions [114] and NS feedback is important for NNSs' communication experience [6]. Therefore, this channel is designed to disclose NNSs' speaking difficulties to NSs. We anticipate that this channel will improve NSs' comprehension of NNSs' circumstances and share communication responsibility [71, 98] through offering constructive feedback, thus reducing the interactional anxiety of NNSs. We examine how this design influences the communication experience of NNSs, especially the impact of the speaking assistant and the mutual understanding channel, while also exploring NSs' perceptions and responses to this intervention.

We conducted a study with 25 NNS-NS pairs (N=50). Each pair completed two distinct collaborative tasks [46], with the task order counterbalanced to mitigate potential ordering effects. The findings revealed that the speaking assistant significantly improved the self-efficacy of NNSs in speaking and reduced their workload during real-time communication. This beneficial effect was particularly pronounced for NNSs with lower L2 proficiency, who reported engaging more with the assistant and experienced the most substantial improvements in their communication. Furthermore, by facilitating the provision of positive and explicit feedback from the NSs, the tool effectively alleviated the anxiety resulting from interpersonal uncertainty, thus providing crucial reassurance and fostering continued engagement from the NNSs. From the NSs' perspective, a prevalent willingness to assume shared communication responsibility was observed, alongside expressed empathy for the challenges faced by their NNS partners.

In summary, this research offers two main contributions to the Computer-Supported Cooperative Work (CSCW) community.

(1) We introduce a tool and empirically demonstrate its effectiveness in mitigating the linguistic and interactional anxiety faced by NNSs, thereby illustrating the tangible potential of AI-mediated communication in real-time collaborative scenarios.
(2) We provide actionable design insights of future AI-mediated communication tools for speaking support, advocating for tool designs that encourage NSs to actively share communication responsibility, and propose novel assistance paradigms.

## 2 Related Work

### 2.1 Linguistic and Interactional Challenges of Non-Native Speakers

*2.1.1 Linguistic Anxiety.* In multilingual communication, NNSs often face significant challenges, including poor comprehension [15], less engagement [69], and inadequate speaking production, where speaking is considered the most challenging [2, 13, 107]. Factors such as limited vocabulary, grammatical difficulties, and struggling to structure thoughts into coherent speech [7, 85] collectively contribute to poor speaking performance. Accompanying these challenges is the NNSs' fear of negative evaluation [66], resulting in linguistic anxiety. These negative impacts further suppress their language performance and hinder NNSs from playing their role in team collaboration [49, 53, 83]. These studies imply the necessity of supporting NNSs' speaking in real-time conversations.

*2.1.2 Interactional Anxiety.* NNSs are also affected by interactional anxiety during real-time conversations, which is caused by uncertainties in conversation [37, 68, 104], such as silence and communication breakdowns [32, 75, 79, 84]. Due to a high cognitive load in multitasking, such as planning upcoming speech or thinking of appropriate words to handle conversation [61, 81, 92], NNSs often experience communication breakdown or pauses [85]. As the human brain naturally seeks predictability and control, a lack of information or uncertainty can lead to a sense of threat and increased alertness, triggering the body's stress response [47]. Levinson [67] further detailed that listeners are inclined to draw strong inferences in such situations: there is either a disruption in the communication channel, or something is socially or interpersonally amiss. In the absence of a clear explanation, recipients of silence may begin to question the interaction itself, wondering if they have said something inappropriate, the interlocutor is unwilling to respond, or if the conversation has broken down altogether.

Taken together, anxiety related to language stems from their lack of confidence about their own abilities and inability to use the language effectively [42], while interactional anxiety arises from opaque conversational dynamics [67].

## 2.2 Existing Methods to Support NNSs

*2.2.1 Support for Comprehension and NNSs' Participation.* Existing studies supporting NNSs in real-time communication are more focused on achieving conversational balance by improving comprehension and creating more opportunities for participation, rather than directly providing support to alleviate linguistic and interactional anxiety in speaking.

To enhance NNSs' comprehension, prior research leverages machine translation [20, 100], automated transcripts [51, 91], and highlighting [43, 50, 90]. Specifically, Pan et al. [91] found that when real-time transcription is provided, the comprehension ability of NNSs can be significantly improved. Gao et al. [43] combined highlighting in translation and indicated that people consider the pairs of messages clearer and less distracting when the keywords in the message are highlighted. Keyword highlighting also improved subjective impressions of the partner and the quality of the collaboration.

Regarding creating speaking opportunities, Li et al. [69] developed an automated agent that monitors conversation dynamics, tracking participants' speaking duration and frequency. The agent intervenes when one speaker, typically an NS, dominates the discussion, creating more opportunities for NNSs to contribute. Similarly, Duan et al. [35] introduced a speech speed monitor that alerts NSs when they speak too quickly, allowing NNSs more time to process information and formulate responses [113]. Another approach by Pan et al. [90] enabled NSs to highlight key content in automated transcripts, redistributing the conversational workload and alleviating communication pressure for NNSs.

*2.2.2 Support for NNS Speaking.* Currently, very few studies provide real-time speaking support. Qin et al. [94] recently proposed an AI-based speaking assistant (AISA), whichallows NNSs to input partially formed ideas when experiencing difficulties, with AI generating contextually appropriate suggestions to support NNSs' speaking in ongoing conversations. While the AISA was found to enhance NNSs' perceived organization of their expressions, it failed to reduce anxiety or perceived workload. In fact, some participants reported that the tool increased interactional anxiety, particularly due to the awkward pauses that occurred while using it, which in turn discouraged their exploration of this tool. Qin et al. [94] attributes this outcome to the system's narrow focus on linguistic inadequacy, overlooking the added interactional anxiety. NNSs in [94] reported their concerns about disrupting the conversational flow and making their NS partner wait for an extended duration.

These findings underscore the complexity of supporting NNSs in speaking: linguistic anxiety and interactional anxiety need to be considered concurrently, and providing solutions that only compensate for speaking ability may backfire.

### 2.3 Research Objectives

From previous literature, we found that speaking support for NNSs should not only reduce pressure in language production, but also focus on interactional anxiety.

To address linguistic anxiety, we use machine translation. It has been widely adopted for multilingual communication support [70, 86, 100], which reduces the cognitive and social burdens associated with L2 usage [4, 12, 27, 99, 102]. For flexibility and responsiveness, this function can be used with voice [97] or typed inputs. When activated, it automatically mutes the user to prevent disruptions to the ongoing conversation.

Regarding interactional anxiety, we introduce a channel aimed at improving mutual understanding between NNSs and NSs [35, 94]. Earlier studies highlight that NSs often lack awareness of NNSs' language difficulties [52, 71, 98]. Duan et al. [35] emphasized that without effective understanding channels, NSs often misinterpret the source of communication difficulties, overestimating NNSs' language abilities and attributing low participation to personality traits like shyness, rather than anxiety or communicative challenges [35, 52]. This suggests the importance of clearly conveying the NNSs' experience, such as linguistic difficulties, to NS. Meanwhile, self-disclosure has been shown to elicit prosocial behaviors, including empathy, from others [106, 114]. We thus posit that signaling NNSs' speaking difficulties and reliance on assistance during real-time conversations may foster greater understanding and supportive behavior from NSs. Additionally, showing NSs' attitudes toward NNSs is also valuable [6, 85] as it can help NNSs interpret social cues, reducing the uncertainty about conversation status, thereby alleviating interactional anxiety [25, 33, 68, 104].

Therefore, we used a common method in prosocial design studies to promote the prosocial behavior of NSs by constructing simple nudges on the interface [77]. When NNSs activate the speaking assistant, NSs receive a fixed notification, *"Please be patient, I am using a translation tool now"*, intended to raise awareness of the challenges NNSs face during speech. Simultaneously, a response panel appears on the NS interface, allowing NSs to react using preset emojis or a custom message to express their attitudes toward the tool's usage (participation is optional).

In this study, we aim to first investigate how our designs impact NNSs' speaking experiences. Therefore, this study poses the following research question:

**RQ1:** How does this tool affect the NNS's speaking experience, including their interactional anxiety, speaking self-efficacy, and workload?

Here, interactional anxiety refers to concerns about conversational interruptions and uncertainty regarding the other person's attitude [32, 37, 68, 75, 79, 104]. Meanwhile, linguistic anxiety is assessed via speaking self-efficacy, grounded in Bandura [8]'s social cognitive theory, which identifies self-efficacy as a key mediator between emotional states such as anxiety and actual behavior. While anxiety measures capture general affective discomfort, self-efficacy more directly reflects the user's confidence in managing the speaking task, making it both actionable and sensitive to intervention.

Furthermore, we seek to understand NNSs' evaluations of the tool's individual components and overall usability with the following research questions:

**RQ2:** How does the design of each component contribute to the NNSs' speaking experience?

- **RQ2.1:** How does the NNS evaluate the speaking assistant?
- **RQ2.2:** How does the NNS evaluate the channel designed for mutual understanding?

In Duan et al. [35]'s study, the introduction of assistance for NNSs, without fostering mutual understanding, resulted in dissatisfaction among NSs. To gain more insights from NSs, we posed the following research questions:

**RQ3:** How does the NS treat and perceive this tool?

- **RQ3.1:** How does the NS view the changes brought about by the tool for the NNS?
- **RQ3.2:** How does the NS respond to the NNS's distress signal, and what are their underlying motivations?

## 3 Method

### 3.1 Overview

In this study, we conducted a mixed-methods study consisting of a within-subjects experiment. We recruited NSs and NNSs of English, pairing each NNS with a unique NS. Each pair completed two collaborative tasks: one with access to the tool and one without. The sequence of tasks and the accessibility of the tool were counterbalanced. After each task, both participants completed a survey regarding their experience working with their partner. The study concluded with interviews to explore NNSs' experiences in greater depth.

### 3.2 Participants

We recruited 25 NSs of English (16 females, 9 males) and 25 NNSs of English (native Chinese speakers) as participants (14 females, 11 males) by posting announcements in local social media communities. Native Chinese speakers constitute approximately 18% of these local communities, while the remainder are predominantly native English speakers. These participants were divided into 25 pairs, each containing one NNS and one NS. We used self-reported English speaking proficiency instead of standardized test scores because our study focuses on subjective experiences, such as anxiety and self-efficacy. Prior work shows that self-perceived proficiency is a significant predictor of foreign language anxiety and often explains greater variance in anxiety than objective proficiency measures, making it a suitable proxy when the focus is on users' subjective experience of speaking [1, 11, 111]. The NNS participants' self-reported English speaking proficiency levels were distributed as follows: 6 below average, 14 average, and 5 good. Regarding education, 1 NNS participant held a high school diploma, 8 had bachelor's degrees, 14 had master's degrees, and 2 held PhDs. On the other hand, 10 NS participants held high school diplomas, and 15 had bachelor's degrees. The mean age of NNS participants was 24.519 ($SD$ = 3.203), and that of the NS participants was 28.957 ($SD$ = 7.010). Participants were compensated at a rate of $13 per hour. The study received approval from the Institutional Review Board at the author's university.

### 3.3 Task Selection

We adopted the Objects Game [46] (see the game interface in Fig. 1) from the Columbia Games Corpus[1], which is commonly used in team collaboration studies. Each round displays identical sets of multiple figures on a square canvas for both players. Anchor figures, marked with a red solid border, cannot be moved, while draggable figures are marked with a gray dashed border. At the start of each round, all figures are arranged in predetermined positions. Anchor figures appear in the same positions for both players, while draggable figures do not. In each round, the NNS acts as the describer and cannot move any figures, while the NS assumes the follower role and can move only their draggable figures. When the round begins, the NNS describes the true positions of their draggable figures to the NS, whose objective is to arrange their draggable figures to match those described by the NNS. Players are unable to see each other's canvas and communicate only through voice and video. Once both players agree on the arrangement, the round score is calculated based on the total distance between the NS's draggable figures and their correct positions. To account for potential variations in screen sizes, distances are calculated as ratios of the game canvas rather than absolute pixel distances. Objective scores in this game could be influenced by various factors, such as body language, individual biases in describing locations, and the NS's questioning style. Therefore, scores are displayed only in the practice round to provide players with a general sense of performance, allowing them to adjust their approaches in formal rounds, rather than serving as an absolute measure of communication quality.

[1]https://www.cs.columbia.edu/speech/games-corpus/

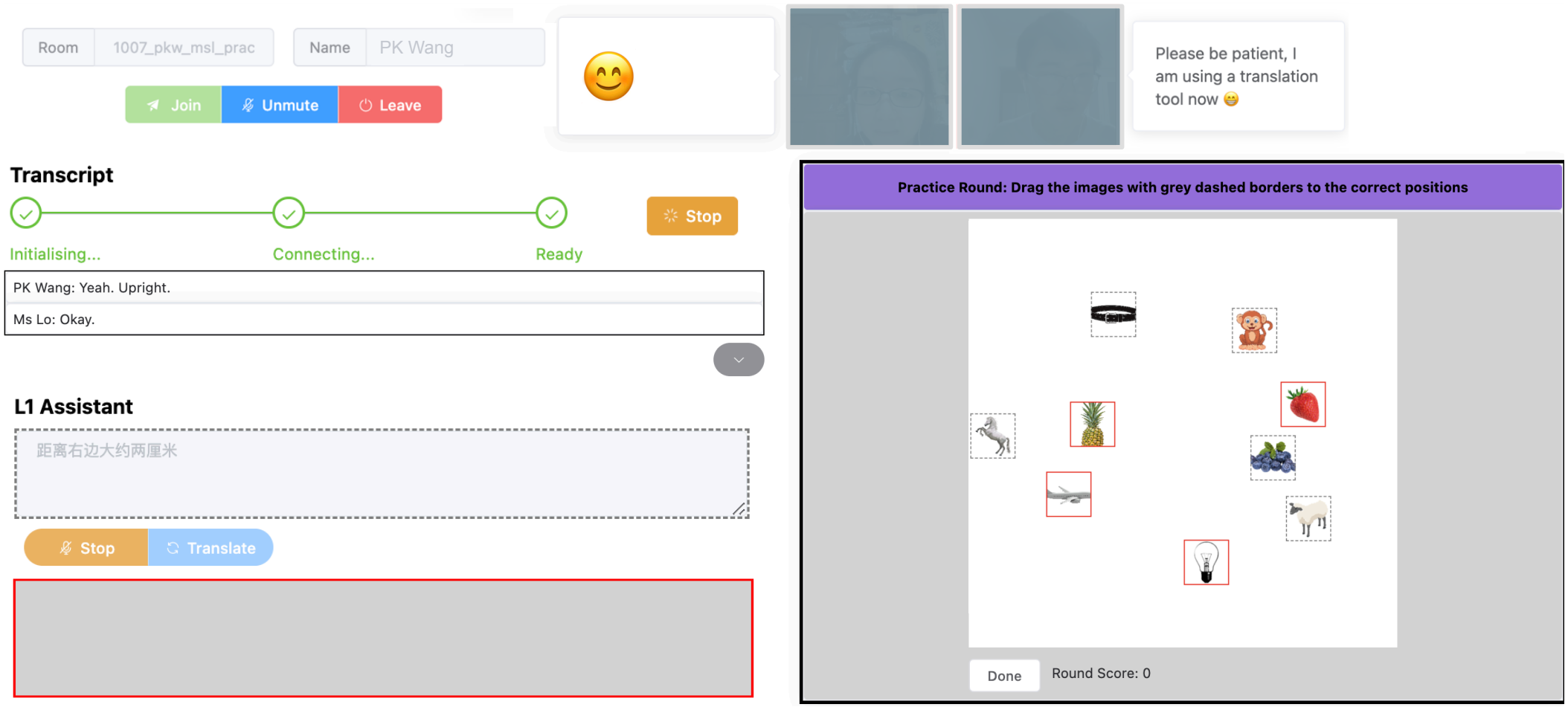

Fig. 1. The experiment interface displaying the tool that integrates with the Objects Game component. The participants' video feeds have been edited in this figure to maintain anonymity. On the left half of the screen, from top to bottom, are the room controls, the folded transcript panel, and the voice-driven translation. In the top right section, the video area displays the video stream while the the pop-ups are channels (1) showing the NNS's help-seeking message and (2) allowing the NS to respond with emojis and custom-message. Lastly, the Objects Game component shows the canvas with anchor figures (red solid borders) and draggable figures (gray dashed borders).

## 3.4 Experiment Platform and Tool Implementation

*3.4.1 Experiment Platform.* The experiment platform (See Fig. 1) was built with the front-end framework Vue.js[2] and the back-end framework Django[3]. Real-time audio and video communication was implemented using Agora[4]. Additionally, automatic speech recognition, supported by Assembly AI[5], was integrated to provide a conversation transcript, aiding NNSs in comprehension and allowing

[2]https://vuejs.org/
[3]https://www.djangoproject.com/
[4]https://www.agora.io/en/
[5]https://www.assemblyai.com/

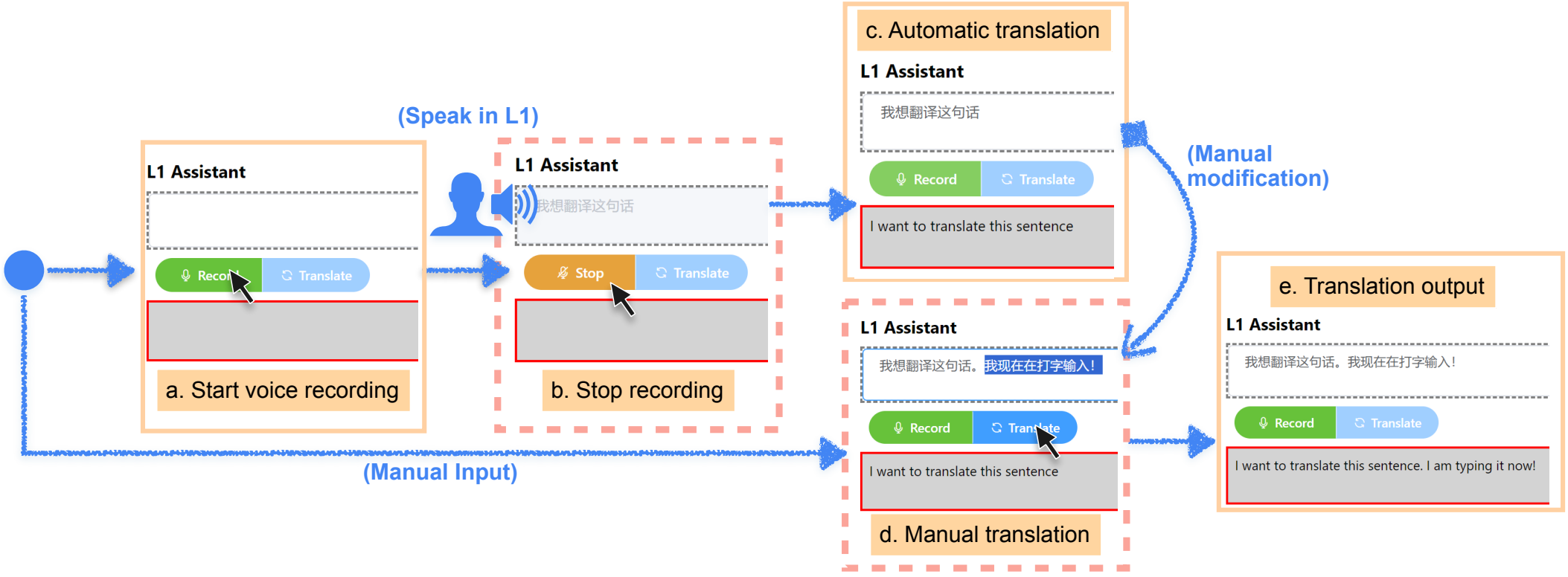

(a) The user journey of an NNS using the speaking assistant. They can choose to use voice or typed input. Starting voice input automatically mutes the NNS for others in the room, and streams the speech to text in the input field. When the recording is stopped, the input will be instantly translated to the L2 and be displayed in the output field. If the NNS prefers to type or edit the current L1 input, they need to click on "Translate" for manual translation. The NNS can refer to the L2 translation output to express their thoughts in the conversation.

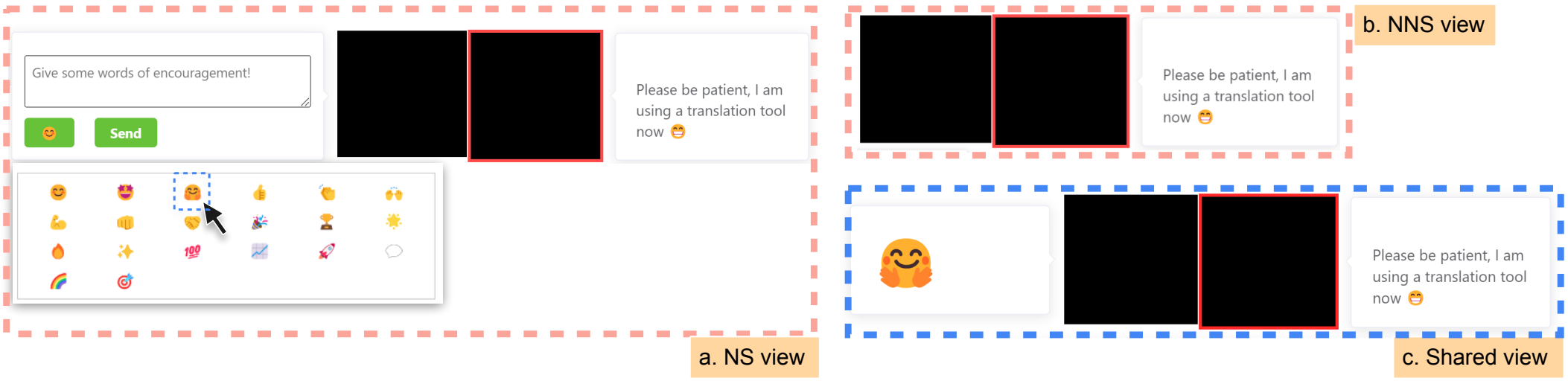

(b) The dynamics of the NNS's and NS's interfaces upon usage of translation and response by NS. When the NNS seeks help to translation, the NNS's video feed will display a flashing red border alongside a pop-up notification "Please be patient, I am using a translation tool now" to notify the NS that the NNS is facing challenges in expressing themselves verbally. Simultaneously, a pop-up response panel will appear only for the NS, allowing them to send a response via emoji or text to convey their attitude towards the NNS's usage of the speaking assistant. The NS feedback will be visible as a pop-up message for both users.

Fig. 2. The user journeys of the (a) speaking assistant and (b) the mutual understanding channel, detailing possible actions and interface changes of the NNS and NS.

them to catch up on missed dialogue [10, 51, 91]. The room controls, located in the upper left corner, allow participants to join and leave the room and toggle their microphones. The transcript panel below it starts displaying the conversation history of the room in real time when participants enter the room. By default, only the two most recent messages are shown to minimize cognitive load for NNSs, with an option to expand the transcript for the full conversation history. The translation component consists of an input field, two buttons, and an output field. In the top right section of the interface, video feeds allow participants to see each other while the pop-ups are channels (1) showing the NNS's help-seeking message and (2) allowing the NS to respond with emojis and a custom message. Lastly, the Objects Game component is comprised of the instructions, canvas, figures, controls, and round score. Section 3.4.2 demonstrates the implementation details and the user journey of this tool.

*3.4.2 Tool Implementation and User Interaction.* First, this tool integrates a voice-driven translation that aids NNSs when encountering speaking barriers. It was implemented by the Google speech-to-text API[6]. To further enhance the robustness and accuracy of the translation, we compared the translation results between the Google Translate API[7] and OpenAI's GPT-4o[8] when the input content had slight errors. These errors could arise due to unclear speech, similar-sounding words, flaws in automatic speech recognition, or mistyped characters. Ultimately, OpenAI GPT-4o performed better as it could be prompted to correct for minor errors, and we chose it to provide our streaming translation service. The prompt we used is:

> Translate the following sentence from {source_language} to {target_language}. Some inputs may be inconsistent due to misspellings, mispronunciations, or errors in speech recognition. Please infer the correct intentions from these errors without any additional information. For numerical or non-{source_language} inputs, respond with the {target_language} equivalent fully spelled out. Do not respond with anything other than the translation content.

As shown in Fig. 2(a), when the NNS uses the translation, they can choose to use voice or typed input. Starting voice input automatically mutes the NNS for others in the room, and streams the speech to text in the input field. When the recording is stopped, the input will be instantly translated to the L2 and be displayed in the output field. If the NNS prefers to type or edit the current L1 input, they need to click on "Translate" for manual translation. The NNS can refer to the L2 translation output to express their thoughts in the conversation.

Fig. 2(b) shows the dynamics of the tool usage. When the NNS uses voice or typing to input the translation component, a pop-up will be displayed beside their video feed with the message *"Please be patient, I am using a translation tool now."* while their video feed will show a continuously flashing red border. This feedback clearly notifies the NS that the NNS is facing challenges in expressing themselves verbally. Meanwhile, a panel will appear only for the NS, allowing them to provide a response to the NNS's usage of the translation. The NS can respond using preset emojis, which effectively express emotions and are easy to understand in communication [19, 45, 60], or type a custom message. Once the NS sends the feedback, it will appear next to the NS's video feed until the next response is sent. This feedback from the NS conveys their attitude towards the NNS's usage of the speaking assistant.

### 3.5 Procedure

First, the NNS, NS, and researchers joined a scheduled Zoom meeting. After signing consent forms, participants were briefed on the experimental platform and procedure in separate breakout rooms. The NNS received a supplementary guide detailing the use of the speaking assistant; they practiced using the tool with a researcher until they felt proficient. Once familiarized, both participants received a link to the experimental platform, logged in, and transitioned from Zoom to the platform. Before starting the formal rounds, participants completed a practice round of Object Game to acclimate to the platform. The official experiment consisted of two rounds: one where the NNS could use the speaking assistant and one where they could not. Task order was counterbalanced. Following each task, participants completed a survey reflecting on their experience. Finally, a 15-20 minute semi-structured interview was conducted to gather detailed feedback, including evaluations of tools' usability, usefulness, and acceptability.

[6]https://cloud.google.com/speech-to-text
[7]https://cloud.google.com/translate
[8]https://openai.com/index/hello-gpt-4o/

### 3.6 Measurements

We utilized tool log data, 7-point Likert surveys, and semi-structured interviews for a comprehensive evaluation.

*3.6.1 Manipulation Check.* We used single-choice questions to assess participants' awareness of the presence of both the speaking assistant and the feedback component, such as *"Was the speaking assistant available in this room?"* The set of questions is included in Appendix A.1.

*3.6.2 Participant Usage Behaviors.* To investigate NNS usage of the speaking assistant, we measured: ***(1) speaking assistant usage count***, tracking how often NNS participants sought help in their L1, and ***(2) speaking assistant input preference***, identifying their choice of input method and rationale in real-time scenarios. For NS participants, we recorded: ***(3)NS response count***, assessing their willingness to provide feedback to NNS, and ***(4) NS feedback preference***, which was assessed by identifying if the NS opted for emojis or personalized text in responses.

*3.6.3 NNS - Speaking Assistant Usability.* To assess the usability of the speaking assistant, we refer to [17] and adapted it into a 7-point Likert scale (1 = Strongly Disagree; 7 = Strongly Agree) with a Cronbach's $\alpha$ of 0.693 , and used statements such as *"I would like to use this speaking assistant frequently when speaking with NSs."* The set of statements are detailed in Appendix A.2.

*3.6.4 NNS - Speaking Assistant Usefulness.* To assess the usefulness of speaking assistant, we refer to [28] and adapted it into a 7-point Likert scale (1 = Strongly Disagree; 7 = Strongly Agree) with a Cronbach's $\alpha$ of 0.794 , and used statements like *"Overall, I think the speaking assistant is very useful for the communication between the NS and me"* The set of statementss are detailed in Appendix A.3.

*3.6.5 NNS - Perception of the NS's Response.* We assessed the possible perceptions of the NNS to the NS's response with four 7-point Likert subscales (Cronbach's $\alpha$ = 0.783), with statements such as *"I thought that the NS's feedback alleviated my anxiety about speaking"*. The set of statements are detailed in Appendix A.4. In the subsequent interview, we asked for more perceptions of NNS about NS's feedback.

*3.6.6 NNS - Speaking Self-Efficacy.* Self-efficacy [9] refers to an individual's belief or confidence in their ability to successfully complete a task or achieve a goal in a specific situation. This indicator is generally considered to have a significant negative correlation with speaking anxiety [78, 89]. Accordingly, we measure changes in speaking self-efficacy to reflect the impact of our proposed method on NNS language anxiety. This indicator can also assess whether the assistance provided by our system effectively supports the overcoming of language-related difficulties, thereby strengthening learners' confidence in their ability to participate in communication. Thus, we modified the speaking self-efficacy 7-point Likert scale (1 = Strongly Disagree; 7 = Strongly Agree) from [42] (Cronbach's $\alpha$ = 0.905) to assess ***(1) performance speaking self-efficacy*** and ***(2) linguistic speaking self-efficacy***. Performance speaking self-efficacy reflects an individual's confidence in completing verbal communication tasks effectively within specific situations. The statements include *"I feel that I participated well in the process of completing the task with the NS"*. Linguistic speaking self-efficacy pertains to an individual's confidence in accurately and appropriately expressing themselves, including the use of precise vocabulary and clearly articulated reasoning and ideas. For example, *"I am confident in using the appropriate vocabulary to express my thoughts clearly in these tasks"*. The set of statements is included in Appendix A.5.

*3.6.7 NNS - Interactional Anxiety.* Communication breakdown and silence [37, 67, 68, 84, 104] can increase feelings of uncertainty among participants and elevate their anxiety levels. To evaluate this interactional anxiety, we developed a 7-point Likert scale (1 = Strongly Disagree; 7 = Strongly

Agree), with a Cronbach's $\alpha$ of 0.851, and used statements like *"I was worried that my speaking would disrupt the flow of the conversation"*. The set of statements is detailed in Appendix A.6.

*3.6.8 NNS - Workload.* We modified and used four dimensions (mental demand, temporal demand, effort, and frustration) of the 7-point NASA-TLX scale [24] to evaluate the workload of NNS when finishing the description tasks (1 = Very Low; 7 = Very High), with a Cronbach's $\alpha$ of 0.767. The statements include "How discouraged, irritated, stressed, or annoyed did you feel?", and the full set is in Appendix A.7.

*3.6.9 NS - Evaluation of NNSs' Speaking Performance.* We adapted the group communication quality scale [72] to assess the speaking performance of NNSs from the perspective of NSs. We retained the indicators of ***clarity*** and ***comfort*** while introducing a ***fluency*** dimension, with a Cronbach's $\alpha$ of 0.860, and statements such as *"I thought I didn't misunderstand what they want to express". The set of statements are included in Appendix A.8.*

*3.6.10 NS - Motivation for Giving Response.* We assessed the possible motivations for the NS to give responses with four 7-point Likert subscales (Cronbach's $\alpha$ = 0.929), using statements like *"I was sure that they encountered language difficulties at that time, so I wanted to offer some of my own support."* The set of statements are included in Appendix A.9. In the subsequent interview, we also asked for additional reasons that NS chose to provide feedback.

*3.6.11 Interviews.* Semi-structured interviews were used to provide in-depth insights into participants' experiences. For NNS participants, we explored their usage experiences, perceptions of the NS partner's feedback, and suggestions for improving the design.

For NS participants, our initial focus was on their perceptions of the communication effectiveness facilitated by our tool. We also explored their views on the acceptability of NNSs receiving assistance through this additional tool. Subsequently, we built upon the motivations of NS to provide feedback in Section 3.6.10. We inquired about their preferences regarding feedback formats and gauged whether the current design and interaction imposed any workload pressure on NS participants. All interview questions are included in Appendix A.10.

All interviews were audio-recorded and transcribed with the participants' consent. We employed thematic analysis [16] to code the transcripts. This process began with two researchers independently coding the transcripts manually. Our coding strategy was primarily inductive, allowing themes to emerge organically from the data, in contrast to a deductive approach that often relies on preconceived notions and intercoder reliability [80]. Throughout the data coding, the researchers ensured that the themes were closely aligned with the research questions. After the initial coding, discussions were held to clarify the rationale behind each code, leading to a consensus through comparison and reflection. This iterative process involved revisiting the transcripts multiple times to ensure that the codes accurately captured the nuances of the participants' experiences. Through this detailed examination, the researchers were able to discern patterns and formulate overarching themes.

# 4 Result

## 4.1 Manipulation Check

Our manipulation checks on the participants' awareness of the presence of both the speaking assistant and the channels for mutual understanding were successful. All NNSs were aware of the presence of the speaking assistant under the corresponding conditions (100%) and would notice the feedback if the NS made a response (100%). Similarly, all NSs reported that they noticed the pop-up notification when NNSs used the speaking assistant (100%) as well as the feedback panel (100%).

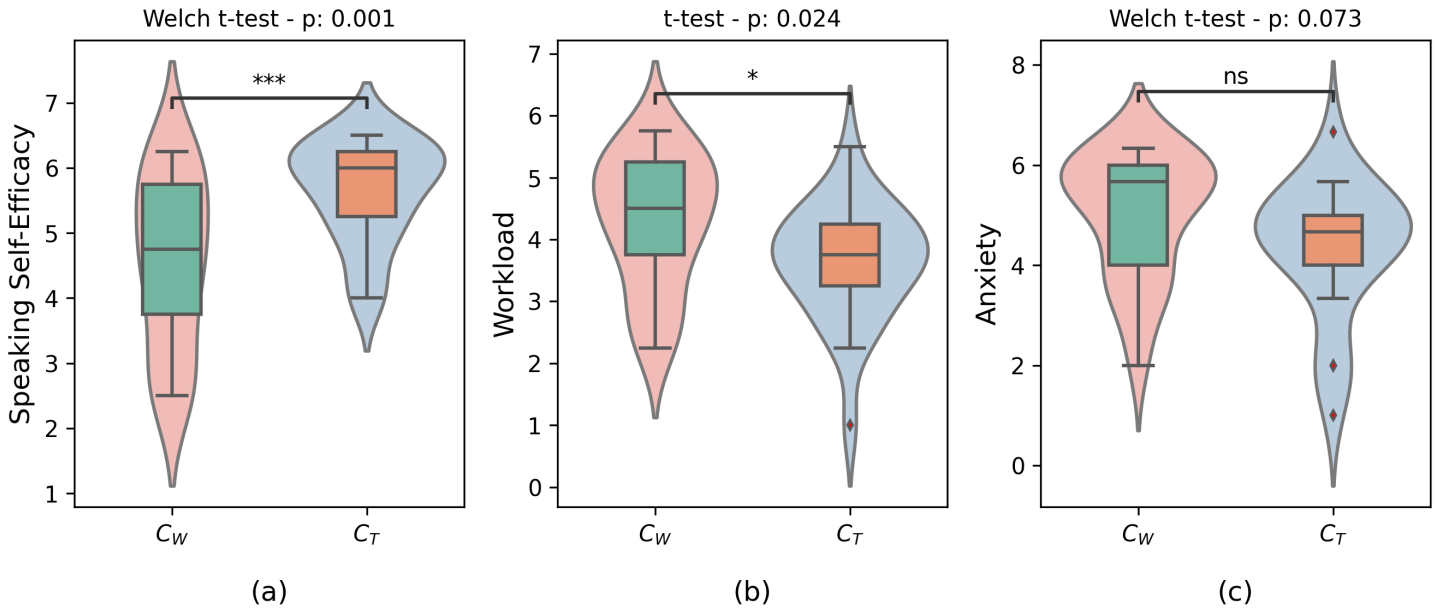

Fig. 3. The comparison on perceptions of speaking self-efficacy, workload, and anxiety between $C_W$ and $C_T$ among **all NNS participants.** The asterisks denote levels of statistical significance, with ns >= 0.05, * representing p < 0.05, ** representing p < 0.01, and *** representing p < 0.001. The violin plots represent the distribution of scores, with overlaid boxplots showing the median and interquartile ranges. In this figure, NNSs showed a significantly higher (a) speaking self-efficacy, significantly lower (b) workload, and a decreasing trend in (c) anxiety, although not significant when our tool was available ($C_T$).

## 4.2 Impacts on the NNS's Speaking Experience (RQ1)

In this section, we mainly use the independent t-test to examine differences in outcomes for NNS participants when the tool is available (denoted as $C_T$) versus when it is unavailable (denoted as $C_W$). The analysis focuses on the impact on NNSs' perceptions, such as speaking self-efficacy, workload, and interactional anxiety. Prior to conducting the t-tests, we assessed the assumptions of normality and homogeneity of variances using the Shapiro-Wilk test and Levene's test, respectively. When assumptions were met, we applied the independent Student's t-test; otherwise, Welch's t-test was used to account for any assumption violations.

*4.2.1 NNSs' Speaking Self-Efficacy.* Overall, the speaking assistant significantly improved NNSs' speaking self-efficacy, with the strongest gains observed among below-average and average proficiency speakers. The Welch's t-test result indicated that there was a significant difference ($t(38.693) = 3.602, p < 0.001, d = 1.019$) in NNS ***speaking self-efficacy*** between $C_T$ and $C_W$ (Fig. 3(a)), due to the availability of the tool. The descriptive data show that the statistic of $C_T$ is $M = 5.640; SD = 0.771$ while that of $C_W$ is $M = 4.540; SD = 1.318$.

Further analysis identified discrepancies within the stratified proficiency levels. In particular, Fig. 4(a) presents the outcomes pertaining to NNS participants who reported their own L2-speaking abilities as *below average*. It shows that although no significant difference in the speaking self-efficacy was observed ($t(10) = 2.037, p = 0.069, d = 1.176$), there was a trend of increasing speaking self-efficacy in $C_T$ ($M = 5.250, SD = 0.612$) while that of participants in $C_W$ was $M = 4.042; SD = 1.317$. For NNS participants who self-reported an ***average*** L2-speaking proficiency, their speaking self-efficacy (see Fig. 5(a)) was significantly improved ($t(26) = 2.834, p = 0.01, d = 1.071$) in $C_T$. The mean for $C_T$ was 5.732 ($SD = 0.717$), compared to 4.625 ($SD = 1.274$) for $C_W$. For the NNS group with *good* speaking proficiency, there was no difference regarding speaking self-efficacy between the two conditions ($t(8) = 1.129, p = 0.291, d = 0.714$), as illustrated in Fig. 6(a).

*4.2.2 Dependence on Speaking Assistant.* NNSs at different L2-speaking proficiency levels exhibited varying degrees of dependence on the speaking assistant (see Fig. 7), reflected by *speaking assistant usage count*. Participants with higher self-reported proficiency used the assistant less frequently.

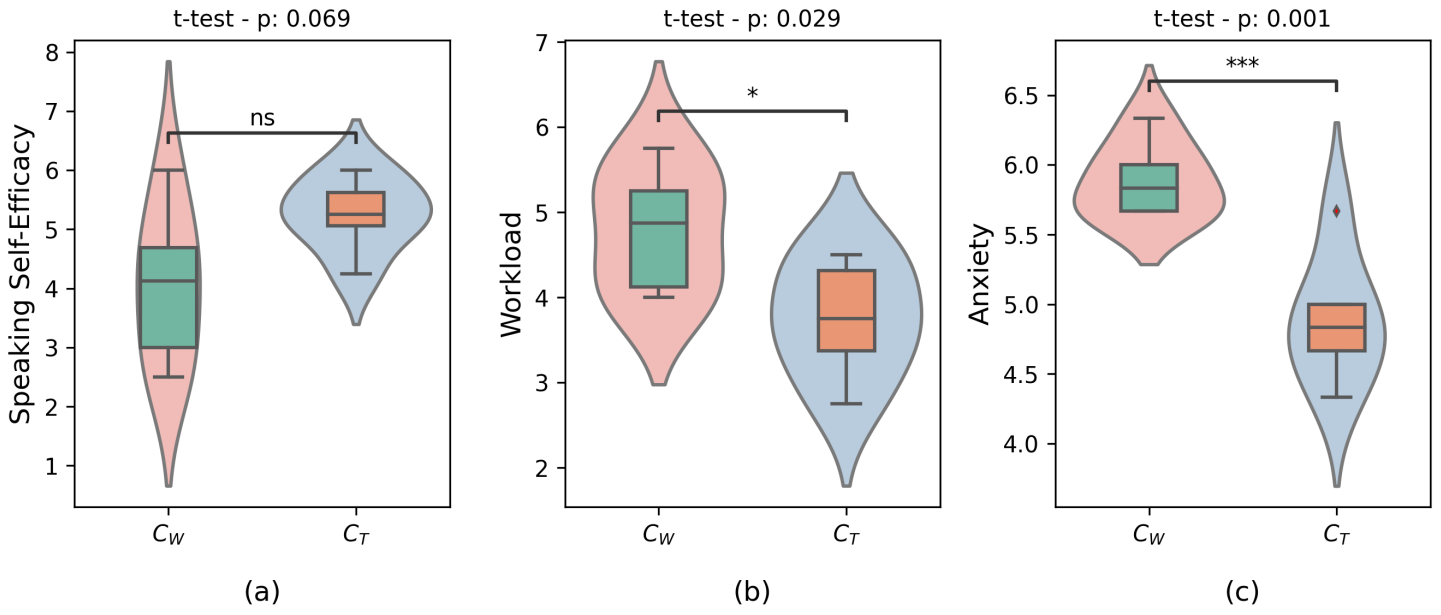

Fig. 4. The comparison on perceptions of speaking self-efficacy, workload, and anxiety between $C_W$ and $C_T$ among NNS participants with ***below average*** speaking proficiency. The asterisks denote levels of statistical significance, with ns >= 0.05, * representing p < 0.05, ** representing p < 0.01, and *** representing p < 0.001. The violin plots represent the distribution of scores, with overlaid boxplots showing the median and interquartile ranges. In this figure, NNS with below average language proficiency showed (a) no significant change in speaking self-efficacy but a higher trend on average, and significantly lower (b) workload and (c) anxiety when our tool was available ($C_T$).

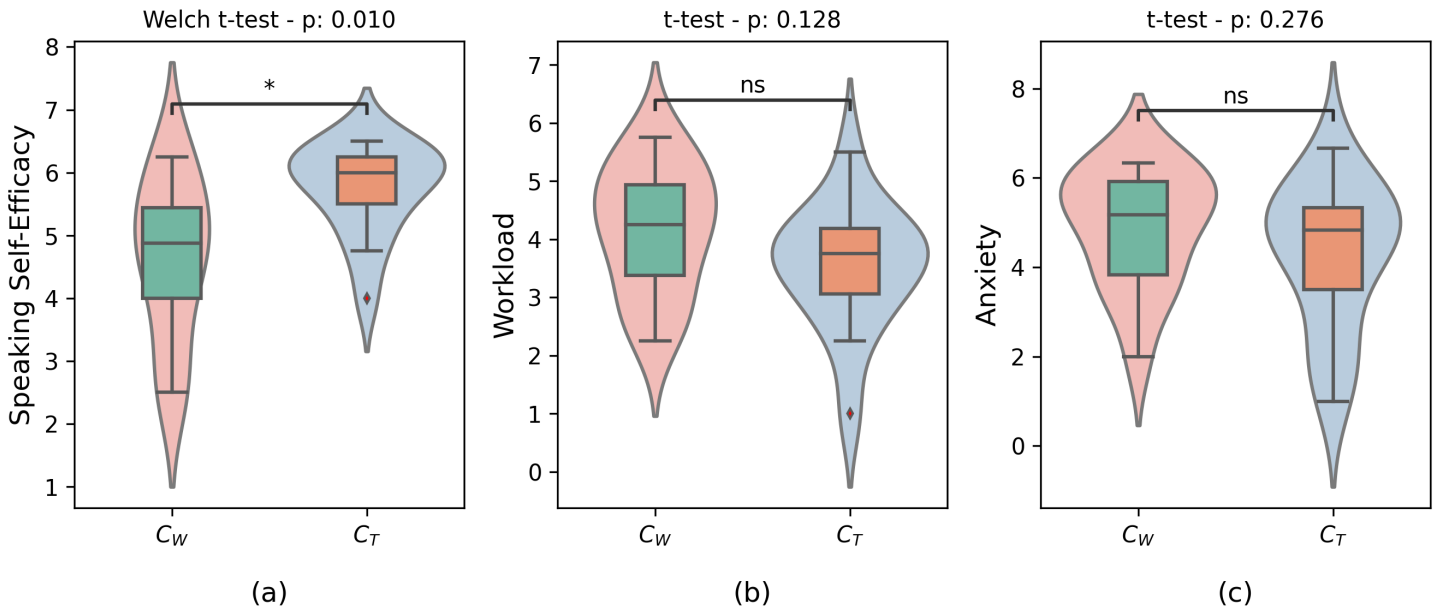

Fig. 5. The comparison on perceptions of speaking self-efficacy, workload, and anxiety between $C_W$ and $C_T$ among NNS participants with ***average*** speaking proficiency. The asterisks denote levels of statistical significance, with ns >= 0.05, * representing p < 0.05, ** representing p < 0.01, and *** representing p < 0.001. The violin plots represent the distribution of scores, with overlaid boxplots showing the median and interquartile ranges. In this figure, NNS with average language proficiency showed a significantly higher (a) speaking self-efficacy. However, the change in (b) workload and (c) anxiety are not significant when our tool was available ($C_T$).

Following Levene's test to ensure variance homogeneity and the Shapiro–Wilk test to confirm normality, the one-way ANOVA showed a significant difference: $F(2, 41) = 6.23, p = 0.004, \eta^2 = 0.233$. The subsequent Tukey post-hoc analysis showed that this difference is notable between the ***below average*** and ***good*** NNS groups ($MD = 4.267, SE = 1.228, t = 3.474, p = 0.003$) as well as between the ***average*** and ***good*** groups ($MD = 2.873, SE = 1.094, t = 2.626, p = 0.032$). NNSs with *below average* L2-speaking proficiency levels used the speaking assistant an average of 8.667 times ($SD = 3.339$), those with *average* proficiency used it 7.273 times ($SD = 2.585$), and NNSs with *good* speaking skills used it only 4.400 times on average ($SD = 2.875$). Regression analysis also indicates that L2-speaking proficiency is a significant predictive factor for the count of speaking assistant usage. Specifically, proficiency negatively predicts input count ($\beta = -0.516, p < 0.001$), explaining approximately 26.6% of the variance ($R^2 = 0.266, F(1, 40) = 14.478, p < 0.001$).

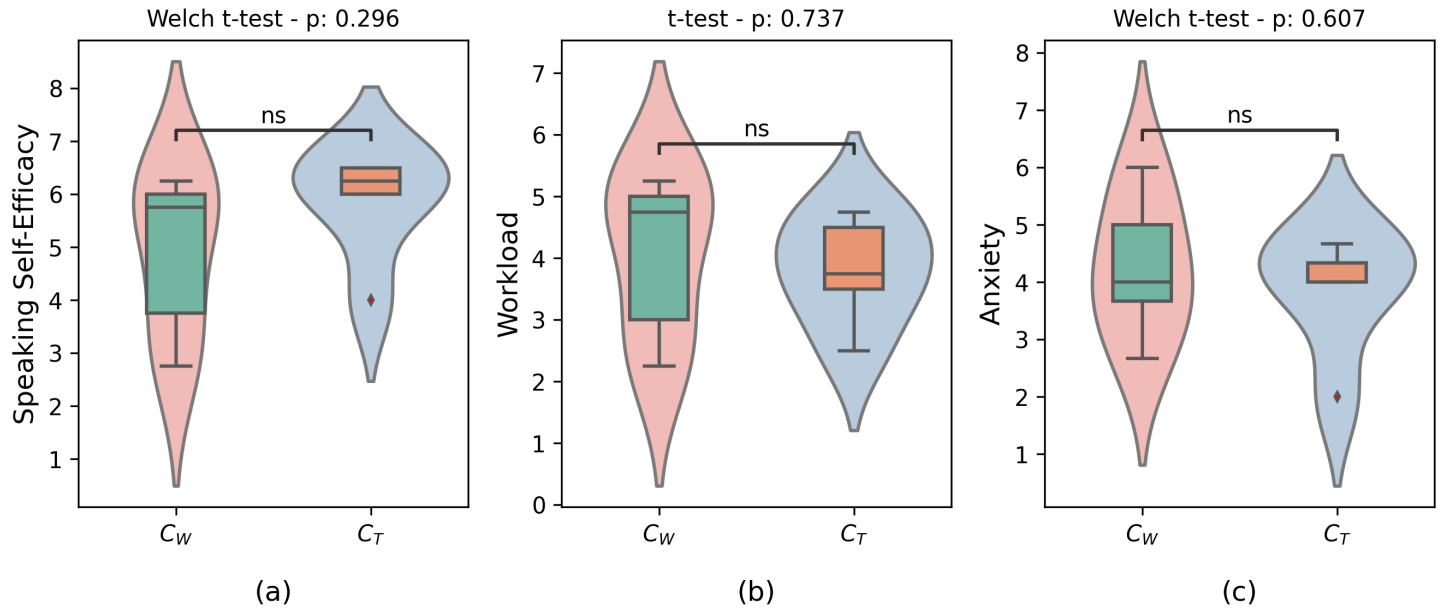

Fig. 6. The comparison on perceptions of speaking self-efficacy, workload, and anxiety between $C_W$ and $C_T$ among NNS participants with ***good*** speaking proficiency. The asterisks denote levels of statistical significance, with ns >= 0.05, * representing p < 0.05, ** representing p < 0.01, and *** representing p < 0.001. The violin plots represent the distribution of scores, with overlaid boxplots showing the median and interquartile ranges. In this figure, NNS with good language proficiency did not show any significant difference in their (a) speaking self-efficacy, (b) workload, or (c) anxiety when our tool was available ($C_T$).

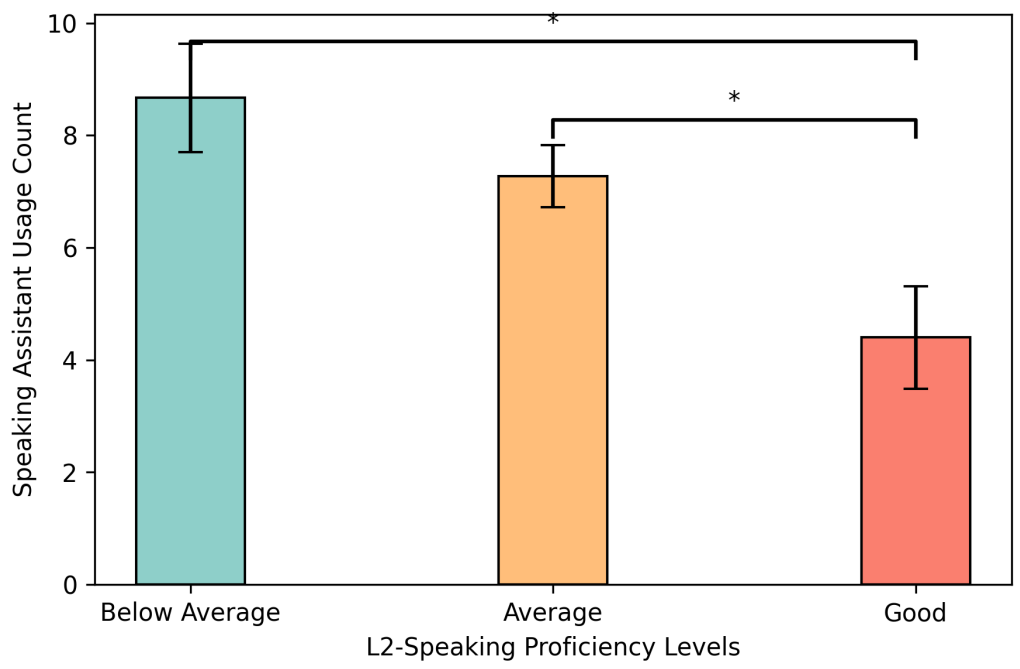

Fig. 7. ***Speaking assistant usage count*** among NNSs, categorized by L2-speaking proficiency level. NNSs with lower proficiency levels exhibited a higher frequency of speaking assistant usage, with a notable decrease in usage as proficiency improved. The significant differences in usage frequency between proficiency levels were confirmed through a one-way ANOVA and subsequent Tukey post-hoc analysis. Error bars represent the standard error (SE) of the mean. The asterisks denote levels of statistical significance, with ns >= 0.05, * representing p < 0.05, ** representing p < 0.01, and *** representing p < 0.001.

*4.2.3 NNSs' Workload.* The speaking assistant reduced perceived speaking workload overall, with statistically significant reductions concentrated among lower-proficiency NNSs. The Welch's t-test (see Fig. 3(b)) illustrates how the usage of our tool can significantly alleviate the ***workload*** of NNSs during interactions with NSs in collaborative tasks ($t(48) = -2.328, p = 0.024, d = -0.658$). When it was accessible, NNSs experienced a mean workload of 3.630 ($SD = 0.930$), compared to a mean of 4.290 ($SD = 1.070$) when it was not accessible.

Examining various L2-speaking proficiency levels, it is evident that our tool has a notable influence ($t(10) = -2.538, p = 0.029, d = -1.466$) on workload for NNSs with ***below average*** proficiency (see Fig. 4(b)). In $C_T$, the statistics stand at ($M = 3.750; SD = 0.689$) compared to $M = 4.792; SD = 0.732$ in $C_W$. For NNSs with *average* ($t(26) = -1.573, p = 0.128, d = -0.595$) and *good* language proficiency ($t(8) = -0.348, p = 0.737, d = -0.220$), the reduction in workload by our tool becomes non-significant (see Fig. 5(b) and Fig. 6(b)).

*4.2.4 NNSs' Interactional Anxiety.* While the speaking assistant did not significantly reduce interactional anxiety overall, a consistent trend towards lower anxiety was observed when the tool was available. To simplify the representation, the following results are uniformly described using ***anxiety***. The Welch's t-test in Fig. 3(c) shows that no significant difference is found ($t(48) = -1.831, p = 0.073, d = -0.518$) due to the tool usage. However, there is a trend of decreased anxiety for NNSs in $C_T$ ($M = 4.333; SD = 1.337$), compared with $C_W$ ($M = 5.000; SD = 1.236$).

Further subgroup analysis showed that there was a significant difference observed in the group of NNSs with ***below average*** language proficiency ($t(10) = 4.617, p < 0.001, d = 2.666$), with NNSs in $C_T$ experiencing lower anxiety ($M = 4.889; SD = 0.455$) while that of NNSs in $C_W$ was $M = 5.889; SD = 0.272$, shown in Fig. 4(c). Conversely, regarding NNSs with *average* ($t(26) = -1.112, p = 0.276, d = -0.420$) and *good* ($t(8) = -0.537, p = 0.606, d = -0.339$) language proficiency, there was no significant difference (see Fig. 5(c) and Fig. 6(c)).

## 4.3 NNSs' Evaluation of the Tool (RQ2)

*4.3.1 Quantitative Evidence of Speaking Assistant's Usability and Usefulness (RQ2.1).* The results indicate that NNS participants who used the speaking assistant found it to have both ***good usability*** ($M = 5.517; SD = 0.750$) and ***usefulness*** ($M = 5.411; SD = 0.738$), while also revealing meaningful variation in expectations around frequency of use and necessity across different speaking contexts and proficiency levels. The detailed breakdown in Table 1 and Table 2 provides more insights.

The descriptive statistics in Table 1 reveal that participants generally viewed the speaking assistant's usability positively, with certain aspects receiving particularly high and consistent ratings. For instance, the statement *"I would imagine that most people would learn to use this tool very quickly"* received the highest mean rating ($M = 6.125$) with the lowest standard deviation ($SD = 0.680$), indicating that ***speaking assistant's features were easy to pick up.*** Similarly, the item *"I thought the speaking assistant was well integrated with the entire tool and easy to use"* also received favorable ratings ($M = 5.625; SD = 1.173$). Another item, *"When I encountered an obstacle in speaking, I immediately thought of using the speaking assistant to help me overcome it"*, also ***reflected high confidence in the speaking assistant's utility*** ($M = 5.583; SD = 1.213$). However, the item *"I would like to use this speaking assistant frequently when speaking with NSs"* had a more moderate mean score ($M = 5.000$) and a relatively high standard deviation ($SD = 1.588$), indicating more varied opinions on its frequent use. This is consistent with the results in Section 4.2.2, possibly suggesting differences in the willingness to use among NNSs with different L2 proficiency levels. These results indicate that NNS participants perceived the speaking assistant as easy to learn, well-integrated into the speaking workflow, and readily accessible when encountering speaking difficulties, though expected frequency of usage varied across individuals.

In terms of ***usefulness*** (Table 2), the descriptive statistics show that participants generally found the speaking assistant to be highly useful in multilingual communication. For instance, the item *"Using speaking assistant made my expression more precise"* received a high mean rating ($M = 5.917$) with a low standard deviation ($SD = 0.654$), suggesting strong agreement among NNS participants. This is also supported by NSs' clarity assessment of NNSs in Section 4.4.1, where NSs considered the ***NNSs' speaking clarity to be improved after using the speaking assistant.*** Crucially, the statement *"Overall, I think the speaking assistant is very useful for the communication between the NS and me"* had an even higher mean rating ($M = 6.083; SD = 0.717$), ***reflecting the broad recognition of its usefulness.*** Conversely, the item *"It would be difficult to communicate without* *the speaking assistant in some real-time multilingual communications with NSs"* had a lower mean rating ($M$ = 4.958) with a larger standard deviation ($SD$ = 1.732), indicating more varied opinions on the necessity of the speaking assistant in specific scenarios. This finding is also consistent with Section 4.2.2, which ***partially supports that NNS with higher speaking proficiency relies significantly less on speaking assistant.*** Moreover, NNS participants rated the speaking assistant positively for enhancing control over speaking ($M = 5.458; SD = 0.884$) and reducing effort in speaking ($M = 5.542; SD = 1.179$). These results can be further reflected in Section 4.2.1 and Section 4.2.3 respectively, suggesting that ***speaking assistant enhances NNS's confidence in speaking and reduces the perceived workload.*** Lastly, *"The speaking assistant made communication smoother and reduced awkward pauses"* received the lowest rating ($M = 4.75; SD = 1.327$), indicating that the speaking assistant showed limited influence on the NNSs' assessment of dialogue fluency, consistent with NSs' evaluation in Section 4.4.1 that ***the speaking assistant did not influence NNSs' speaking fluency a lot.*** Overall, the speaking assistant was evaluated as highly useful for improving expression precision, confidence, and perceived control in multilingual communication, while offering more limited and uneven benefits for conversational fluency.

*4.3.2 Qualitative Evidence of Speaking Assistant's Usability and Usefulness (RQ2.1).* In the qualitative data, NNS participants reported scenarios in which they found the speaking assistant helpful, sharing specific feelings about its usage. Several NS participants also provided insights about the speaking assistant's effectiveness based on observations of their partner. However, a few NNS participants described situations in which they might struggle to use the speaking assistant or even consider discontinuing its use, highlighting areas for potential refinement in the tool's functionality.

*Speaking Assistant Helps in Expression and Alleviating NNS anxiety.* 16 NNS participants explicitly stated that they would use the speaking assistant when encountering ***vocabulary that was difficult to describe*** or ***concepts that were hard to express***. For example, NNS-15 mentioned that the speaking assistant *"reduced [their] thinking time, allowed [them] to answer the question faster, and reduced [their] embarrassment."* Furthermore, it alleviated NNS-15's anxiety as they highlighted its helpfulness *"when [they are] panicking and don't know how to answer the other person's question"*.

Similarly, ***NS participants expressed positive attitudes toward the speaking assistant.*** They noticed the pop-up messages when their NNS partners used the speaking assistant, and a majority (N = 18) found it beneficial. For instance, NS-13 highlighted the support it provided, recalling the embarrassment of her NNS partner when the speaking assistant was absent: *"Without it, my [NNS] partner was flustered and awkward because she could not describe the objects in the game."* Additionally, four NSs observed ***increased confidence and clarity*** in NNS communication following the use of the speaking assistant. For instance, NS-19 noted, *"I believe that it is useful as my partner did manage to convey her thoughts well to me."* NS-21 believed that the speaking assistant significantly enhanced the NNS's ability to express themselves coherently and facilitated better understanding: *"It helps my NNS partner to describe the item ... in English from their native language. The NNS is able to form more coherent and grammatically correct sentences in English ... there is less confusion (and perhaps frustration) ... when trying to comprehend".*

*Situations where the Speaking Assistant is Unwelcome.* Despite the benefits of the speaking assistant, several NNS participants preferred to ***avoid using external assistance whenever they had alternative expression strategies.*** For example, NNS-23 felt that it was only necessary to use the speaking assistant for specific vocabulary barriers, as minor grammatical inaccuracies did not hinder NS comprehension: *"if I can say these words, I may not use [it] because native speakers can understand".* Others also explained that they would only use the assistant when they lacked alternative strategies to express themselves: *"I didn't use it. Because I can find another way to describe* *it. There are a lot of alternative words (NNS-25)." "In the situation that I can find another way to express like using body language, I will give up using it (NNS-13)."*

Two NNS participants also ***raised concerns about the tool's appropriateness in certain settings***, such as interviews or work meetings. NNS-7 stressed, *"if we have meetings with tutors, that could be awkward."* NNS-8 echoed this sentiment: *"Sometimes I can't speak or input voice commands, such as during an interview, I would give up using it."*

*Suggestions for Improvement.* Lastly, a few NNS participants suggested enhancements to improve the speaking assistant's usability. They proposed features like ***direct audio output for complex translations*** or an ***option to share the translation with NSs via text***, as these would help address pronunciation challenges. These NNSs expressed, *"I hope ... the translated text can be sent out directly because ... I do not know how to pronounce them accurately (NNS-23)"*. Another agreed, stating, *"It could be better if it could speak the English translation rather than just displaying the words (NNS-7)."*

In summary, while the speaking assistant was generally well-received by NNS participants for its usability and usefulness in overcoming language barriers, some participants noted alternative expression strategies, expressed concerns about its use in formal settings, and recommended additional features for improvement. The assistant was deemed valuable but could benefit from further refinements to better support diverse communication contexts.

*4.3.3 Preference of Using Speaking Assistant.* Between the two input methods for the speaking assistant: voice input and typed input. The purely typed input was used in 22 out of 151 cases. As depicted in Fig. 8(a) a majority of inputs (129 out of 151) were initiated using voice, and a significant portion of these (75 out of 129) were subsequently modified through typing. When considering the input length (see Fig. 8(b)), there is a significant preference for using voice input over typed input for longer texts by NNSs, as indicated by $t(46.368) = 2.886, p = 0.006, d = 0.528$. Specifically, the mean length for voice input is 13.639 ($SD = 15.702$), while the mean length for typed input is 6.909 ($SD = 8.880$).

Our interview data supports this finding, showing that most NNS participants prefer voice input as their primary method for its speed and simplicity. Specifically, 12 NNSs stated that voice input meets the need for real-time results and efficiency, only opting for typed input when speech recognition issues arise with their L1. For example, *"I use L1 via voice input because it's quicker and saves a lot of time. But voice if there is an error in recognition, I will use typed input to correct it (NNS-15)"*. Likewise, NNS-17 noted, *"I use the L1 assistance through voice input. Since in this way, it's quite convenient for me to get the results quickly and timely."*

However, six participants explicitly preferred typed input, citing the need to avoid disruptions in their speech and the difficulty of switching languages mid-conversation. NNS-3 explained, *"Using the voice input might disrupt my logical flow ... It is hard to me to switch to [oral] Chinese then switch back immediately."* Others echoed similar concerns, such as *"I don't want to suddenly change languages when speaking to others (NNS-21)"*, while another emphasized that language switching *"breaks the immersion of speaking English (NNS-17)"*.

Additionally, some participants found typed input better suited for simpler tasks and more aligned with their habits. One participant explained, *"I use typed input because it's faster when I'm only searching for a few words rather than a whole sentence (NNS-10)"*. Another participant remarked, *"Using typed input is a subconscious choice. When we chat online, we tend to use typing more than voice input (NNS-13)"*.

Overall, while most NNS participants favor voice input for its efficiency, some prefer typed input to maintain conversation flow and out of personal habit. These preferences reflect the diverse needs and usage patterns of NNSs when interacting with the speaking assistant.

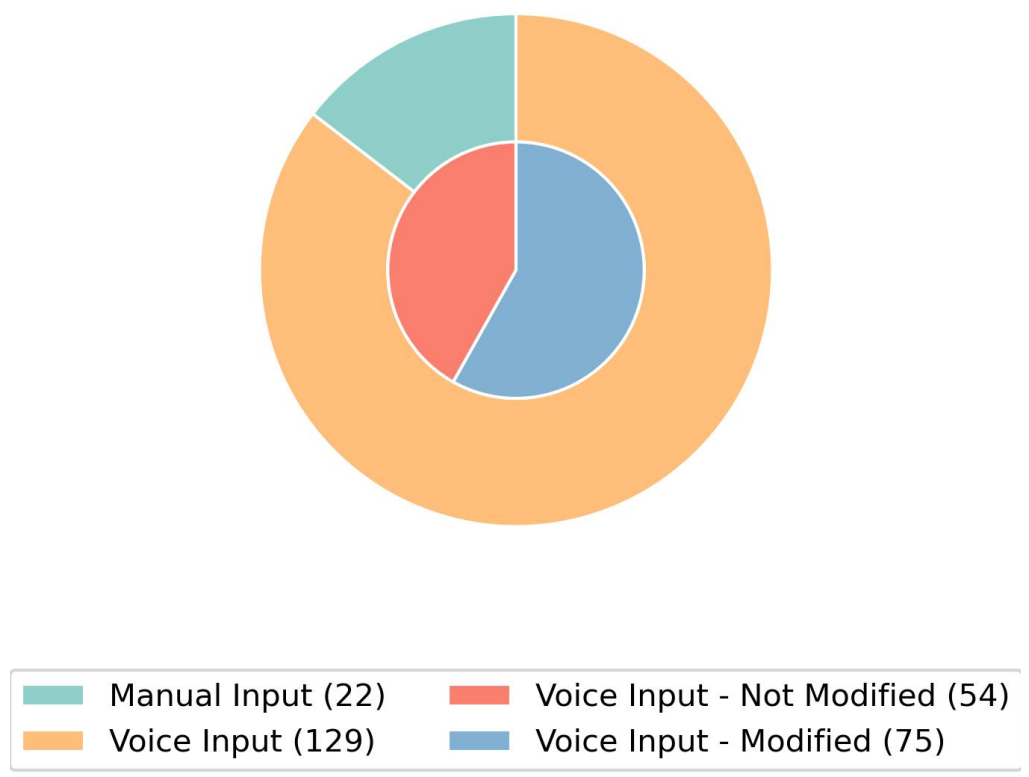

(a) Nested pie chart illustrating the distribution of speaking assistant input methods. NNSs' input can be divided in three ways: merely using manual input (N = 22), using voice input without modification (N = 54), and using voice input followed by manual adjustment (N = 75).

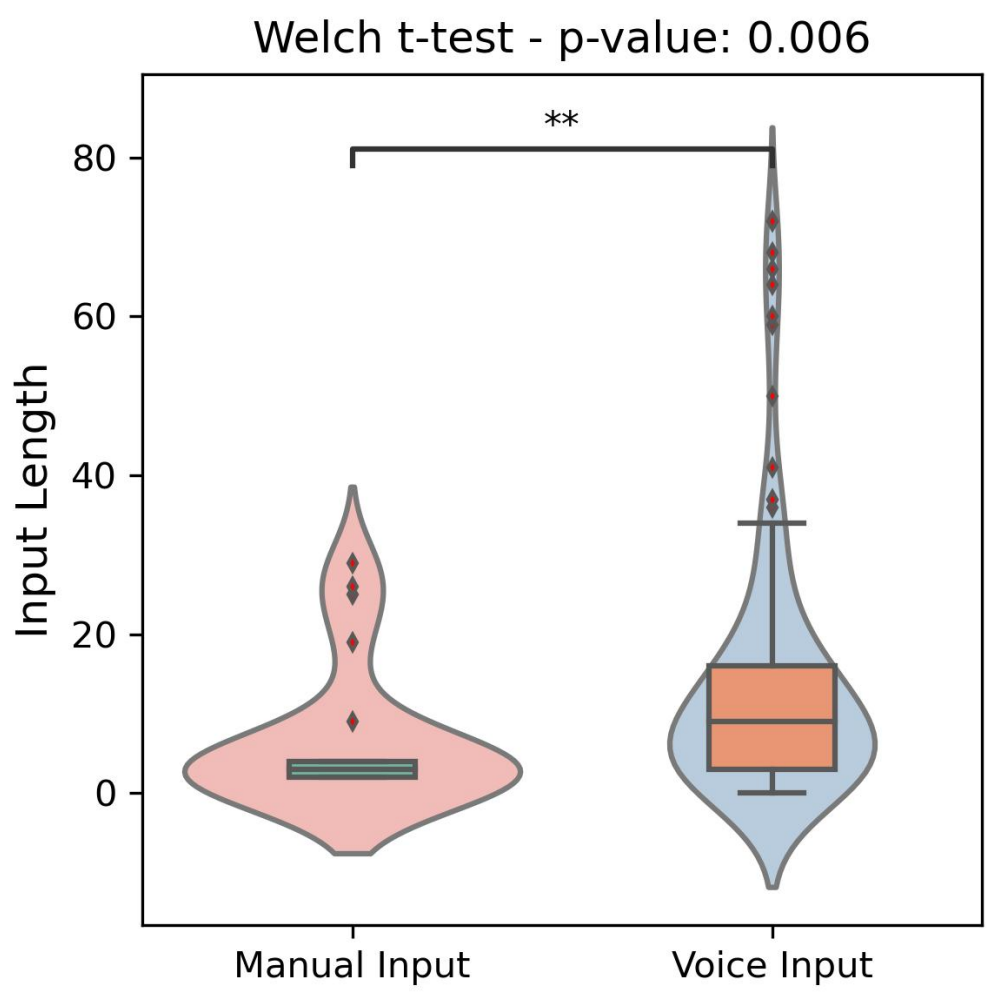

(b) Comparison of input lengths between voice input and typed input. The asterisks denote levels of statistical significance, with ns >= 0.05, * representing p < 0.05, ** representing p < 0.01, and *** representing p < 0.001. The violin plots represent the distribution of scores, with overlaid boxplots showing the median and interquartile ranges. In this figure, the voice input is significantly preferred to meet longer input requirements.

Fig. 8. Descriptive statistics of NNS participants' input behavior when using the speaking assistant for speaking support. (a) represents the distribution of input methods; (b) represents the distribution of input lengths for each input method.

*4.3.4 Quantitative Evaluation of NSs' Responses (RQ2.2).* Overall, NNS rates their NS partner's feedback positively ($M = 5.792; SD = 0.759$). Table 3 presents detailed statistics and suggests that NS feedback is generally well-received by the NNS. For instance, the statement *"I thought that the feedback content from the native speaker alleviated my anxiety about speaking"* received a mean score of 5.611 ($SD = 0.979$), indicating that the NNS participants largely agreed with this sentiment. Similarly, the feedback was perceived as an indication that the NS was willing to wait for them to use the speaking assistant ($M = 5.722; SD = 1.018$). Additionally, NNS participants reported a generally positive feeling toward the NS feedback ($M = 5.889; SD = 1.023$). Notably, the highest mean score ($M = 5.944; SD = 0.873$) was given to the statement that NS feedback helped them better understand the NS's feelings, potentially eliminating some doubts and negative emotions. These findings suggest that NS feedback could reduce anxiety and promote positive attitudes in communication.

*4.3.5 Qualitative Evaluation of NSs' Responses (RQ2.2).* Some NNS participants subsequently elaborated on their experiences with NS feedback and its role in reducing their anxiety. For example, NNS-15 described that *"When I first started to use it, I was nervous to see NS, I was worried that I couldn't say it, my pronunciation wasn't standard, or I couldn't understand NS when I said it, but after using the [speaking assistant], I could express myself very quickly and then translate it over to speak English ... I slowly gained confidence when I saw NS's feedback, I was more willing to express my thoughts and ideas and the positive feedback gave me more courage to speak."* Similarly, NNS-13 stated, *"NS feedback greatly reduces the pressure of expressing myself in English. It reassures me that they are willing to communicate even if I use tools"*. NNS-5 expressed surprise at receiving feedback: *"I didn't expect a response from NS, but it relieved my pressure because it showed the partner's patience and understanding."* Notably, many NNS participants displayed a common behavior: despite the pop-up notification that they were using the speaking assistant, they still verbally reminded the NS, often saying, *"Sorry, I need to use this tool."* These findings underscore the significance of NS feedback in building NNS confidence and alleviating anxiety in multilingual communication. NNS participants are more likely to expect acknowledgment and approval from NS, especially when relying on additional support.

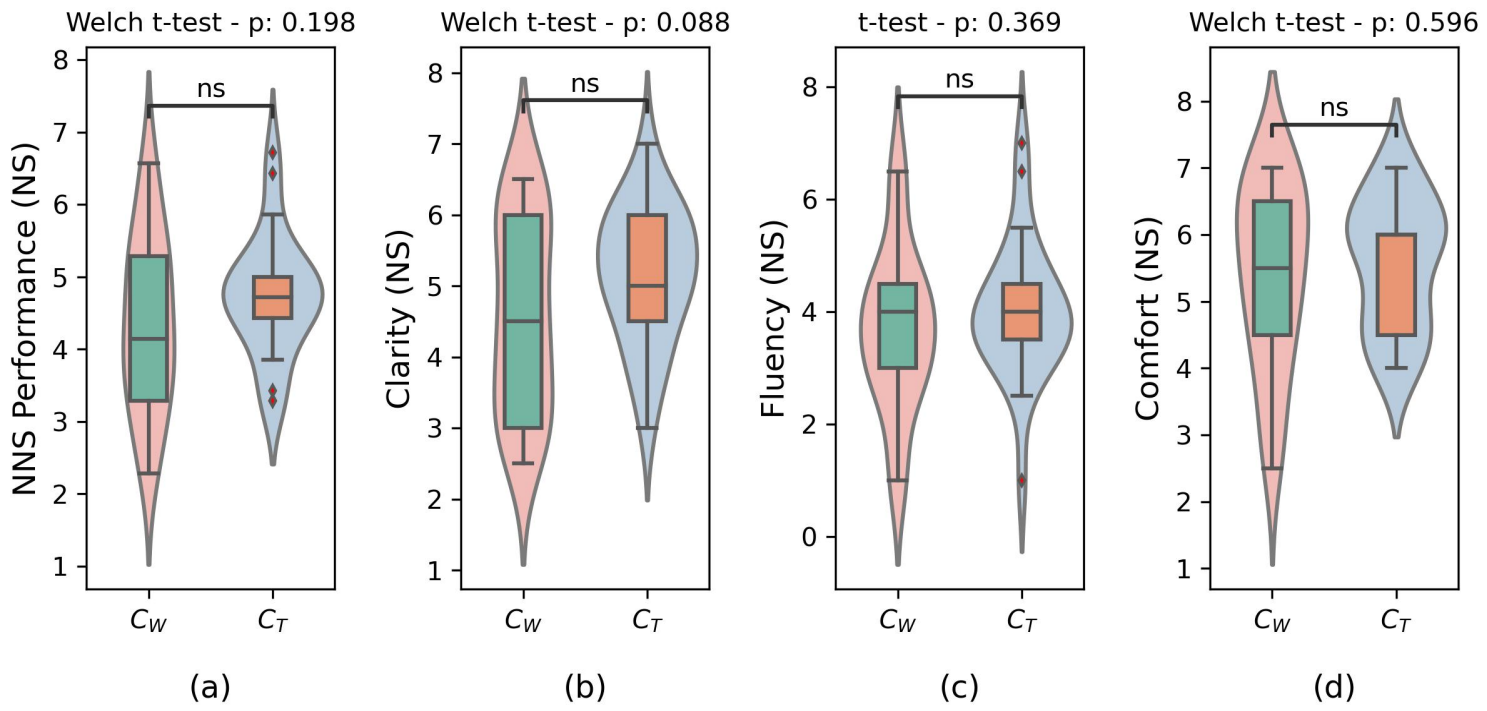

Fig. 9. The (a) NNS speaking performance was evaluated by their NS partner, including three aspects: (b) clarity, (c) fluency, and (d) comfort. The (e) object game score quantitatively measured the collaborative performance between NNS and the NS partner. The asterisks denote levels of statistical significance, with ns >= 0.05, * representing p < 0.05, ** representing p < 0.01, and *** representing p < 0.001. The violin plots represent the distribution of scores, with overlaid boxplots showing the median and interquartile ranges.

In conclusion, NNSs generally perceive NS feedback to be effective in alleviating anxiety and uncertainty about NSs' attitudes, as well as building confidence to speak and use the speaking assistant.

## 4.4 NSs' Evaluation of the Tool (RQ3)

*4.4.1 Effectiveness on NNSs' Speaking Performance (RQ3.1).* Fig. 9(a) shows that when the accessibility of the tool changes, there is no significant improvement in NNS speaking performance ($t(42.734) = 1.309, p = 0.198, d = 0.370$) from NSs' perspective. Regarding various dimensions of the evaluation (Fig. 9(b) - (d)). The NNSs' ***clarity*** shows a trend of improvement ($t(48) = 1.746, p = 0.087, d = 0.494$) although it still holds no significance. The descriptive statistic in $C_T$ is $M = 5.160; SD = 0.965$ while that of $C_W$ is $M = 4.580; SD = 1.352$. No significance is found regarding *fluency* and *comfort*.

*4.4.2 Acceptability about NNSs' Speaking Assistant Usage (RQ3.1).* A majority of NS participants (N = 21) consider the speaking assistant acceptable for real-time communication when NNSs use it to aid verbal interactions. Among them, 11 NSs felt that the ***speaking assistant could enhance communication quality, assist the NNS in minimizing awkward pauses, simplify expressions, and boost confidence; the time required to use it was considered insignificant*** in comparison to these advantages. For instance, *"I think the duration ... is acceptable. The only impact ... was a slower conversation but [it] is alright if I can understand what the NNS was trying to convey (NS-21)." "[The delay] is comparable to the amount of time taken to think about what to say without* *the assistance tool as well. It helps our communication be clearer (NS-19)."* Four NSs further reported that their NNS counterparts seamlessly and naturally employed the speaking assistant to engage in the conversation to the extent that they barely noticed its usage: *"it's fairly fast enough for them to get the answer they need (NS-20)."*

This suggests that the time delay in delivering real-time support via the speaking assistant was not a significant issue for NSs. Instead, NSs place a greater emphasis on the clarity of NNSs' expressions and the quality of communication.

*4.4.3 Preferences of Responses (RQ3.2).* A total of 13 NS participants provided feedback via the pop-up panel when their NNS partner used the speaking assistant, collectively offering 28 responses, of which 26 were positive emojis and 2 were text messages. Then, qualitative findings indicated a strong preference for emojis over text as they were ***faster and easier to communicate***, enabling NSs to offer encouragement without disrupting their NNS partners' focus on the speaking assistant. As NS-15 mentioned, *"Emojis are easier ... Text is weird to me as I will need to think."* Another highlighted efficiency, stating, *"Emojis are much easier to interpret when the timeframe is tight ... people process faster seeing an illustration than reading words (NS-23)."* Furthermore, NSs felt that ***emojis provided a "common ground"*** (NS-9) across language barriers, reducing the need for clarification. In contrast, ***typing text required additional effort and risked distracting NNS participants***. For instance, NS-24 commented, *"[Text] might distract the NNS from organizing their thoughts."*

In summary, NS participants primarily used emojis over text when providing feedback during their NNS partners' use of the speaking assistant. Key reasons included the speed, simplicity, and non-intrusive nature of emojis, which allowed NSs to show encouragement without interrupting their partners' concentration. Additionally, emojis served as a universally understood form of communication, minimizing potential misunderstandings.

*4.4.4 Reasons to Give or Withhold Responses (RQ3.2).* Table 4 presents the intrinsic motivations of NS participants who provided feedback, revealing that a sense of responsibility was the strongest motivation, with the highest mean value ($M = 5.846; SD = 1.345$). Other significant motivators included the desire to alleviate NNSs' anxiety ($M = 5.615; SD = 1.446$) and an awareness of language challenges faced by NNSs ($M = 5.538; SD = 1.613$). While a willingness to calm NNSs down also influenced NSs' decision to give feedback, this was relatively less prominent ($M = 5.308; SD = 1.494$).

Two NS participants expressed empathy, drawing from personal language-learning experiences. NS-23 shared, *"I failed every english exam up till [polytechnic]. Hence, ... I don't wish for [anyone] to feel shame or stressed [about] language ever."* Similarly, bilingual NS-7 performed a role reversal and wanted to *"[let] them know it is okay to take a while to think about how to say it ... I can imagine how it would be like if it was flipped".*

However, some NS participants opted not to provide feedback via the available channels. A common reason was the ***concern that feedback might interfere with the NNS's concentration.*** For instance, NS-13 stated, *"I did not want to distract my partner because she was typing and focusing on translating."* Three NSs felt that ***real-time body language and facial expressions were sufficient for conveying support***, with NS-6 commenting, *"...body language and facial expressions ... [are] sufficient to express [our] attitudes without the need to use additional emojis or text..."*. NS-18 echoed this sentiment: *"...verbal communication and body language ... was enough to demonstrate patience and encouragement."* Another three NSs simply ***"didn't see the need to use the feedback channels".***

In summary, except for a very small number of NSs who still believe it is unnecessary to provide feedback, other NSs seem to recognize the need to take on more responsibility in communication. Therefore, some NSs directly expressed positive emotions, while others considered to avoid disturbing NNSs by not giving a response or providing feedback through other non-invasive methods.

## 5 Discussion

This study initially demonstrates an improvement in speaking self-efficacy and a reduction in perceived workload among NNSs who use the system. Furthermore, our findings reveal the differential effects of the assistant on NNSs with varying levels of L2 proficiency. Across conditions where the system was available, we observed increased NS awareness of NNSs' communicative challenges and more empathetic interaction patterns, alongside reduced interactional anxiety among NNSs. These patterns are consistent with the presence of system features that surfaced NNS difficulty to NSs, though the study does not isolate the effects of individual components. However, it is noted that the system showed limited direct improvement in NSs' observable speaking performance metrics. The subsequent subsections will delve into the key findings of our study and culminate in actionable design recommendations for future research and development in AI-mediated multilingual communication.

### 5.1 Effectiveness of the System and the Phenomenon of Stratification

The results show an increasing trend of speaking self-efficacy, decreasing anxiety and workload when the system was available, indicating its potential to support NNSs in overcoming linguistic barriers [5, 30, 109] and interactional anxiety, achieving the initial design goals. Because our study compared the system against a no-assistance condition, we interpret these effects at the system level and do not attribute outcomes to individual components. The study, together with Qin et al. [94]'s AISA, enriches the research of AI-mediated communication on supporting NNSs' real-time speaking.

Quantitative and usage data consistently revealed a stratification effect: the system was most beneficial for NNSs with *below average* to *average* L2 speaking proficiency. These individuals exhibited more pronounced improvements in self-efficacy and greater reductions in workload and anxiety, and they also engaged with the system more frequently. In contrast, NNSs with higher proficiency experienced minimal changes in these perceived outcomes and engaged with the tool less often. This disparity likely reflects not only objective differences in linguistic needs, but also variations in help-seeking behaviors, which are often intertwined with self-esteem and identity [108]. Prior research indicates that individuals with high self-efficacy and proficiency tend to rely less on external assistance, sometimes viewing help-seeking as an admission of inadequacy that could threaten their self-perception of competence and autonomy [18, 29, 58, 59]. Consequently, some more competent NNSs might initially opt for familiar communication strategies [3, 34] over the speaking assistant (see Section 4.3.1).

Conversely, NNSs with lower L2 speaking proficiency used the system more frequently, which can introduce a risk of reliance on external assistance. In goal-oriented real-time conversations, users experience a higher multitasking and cognitive load [92] and may prioritize task completion by quickly referencing the system's output without fully processing and learning from it. Prior studies suggest that when users experience greater difficulty, they are more inclined to depend on AI assistance [93], which can undermine their L2 development and language competence [57].

These findings imply that speaking support should be adaptive to proficiency and user preference. For lower-proficiency NNSs, tools may offer richer scaffolding but also supply learning opportunities such as reflective comparison between suggested and intended expressions. For higher-proficiency NNSs, support should be less intrusive and more optional, complementing rather than replacing their existing strategies.

### 5.2 The Essentiality and Concerns of Facilitating Mutual Understanding between NNSs and NSs

The integrated channel for mutual understanding appears to have achieved its primary objectives. Firstly, automated help-seeking messages is consistent with alerting NSs to their communication responsibilities and fostering empathy for the difficulties experienced by NNSs. This illustrates the applicability of principles related to triggering prosocial behavior through strategic self-exposure to vulnerability [114]. In many prior studies [23, 98], it has been observed that NSs often do not readily assume an equitable share of communication responsibility, often attributing breakdowns or difficulties solely to the linguistic limitations of the NNS. Ashtari [6] pertinently emphasized that effective communication is a collaborative achievement, referencing Schegloff [98]'s assertion that *"conversation is always collaboratively achieved"* [23], a concept also known as the principle of mutual responsibility. In contrast, the majority of NSs in our study actively provided supportive feedback, citing a sense of responsibility as their primary motivator. This suggests that a considerable portion of NSs acknowledged shared communication responsibility. This positive shift appear to be aligned with the presence of the notification message, which explicitly informed NSs about the NNSs' speaking challenges and their need to utilize the speaking assistant, displaying the effectiveness of self-disclosure in encouraging empathy and subsequent prosocial behaviors [106, 114]. Notably, several NSs who did not send an explicit response via the panel still reported feeling empathy, choosing instead to offer support in more subtle ways or refraining from overt feedback to avoid potentially disrupting the NNS's focus; only a minority of NSs deemed feedback unnecessary.

Concurrently, NSs generally perceived the speaking assistant as a necessary and valuable tool for helping NNSs articulate their thoughts with precision. They expressed a willingness to accommodate the slight interactional delays introduced by the assistant's use in return for improved overall communication clarity. This finding contrasts with that of Duan et al. [35], in which an agent was designed to interrupt NSs to create speaking opportunities for NSs, yet failed to adequately establish a shared understanding of the rationale behind the intervention, which led to perceptions of unfairness among NSs. This underscores a critical point: while NNSs undeniably face disadvantages in spoken L2 communication and NSs should bear a portion of the communicative responsibility [98], it is imperative to ensure that NSs are fully cognizant of the rationale and necessity when assistive technologies are introduced to support NNSs.

Furthermore, the feedback provided by NSs via the mutual understanding channel was perceived as contributing to a friendly communicative atmosphere for NNSs, offering them reassurance and thereby alleviating their anxiety. Specifically, these responses conveyed a positive attitude from the NSs, which directly addressed the concerns of the NNSs about interpersonal uncertainty [37, 68, 104], such as the disposition of their NS partner or the interpretation of their contributions (see Section 4.3.5). The use of emojis served as a lightweight means for NSs to convey emotional grounding [21, 26, 45, 54, 110], efficiently conveying positive signals.

While the mutual understanding channel effectively fosters empathy, shared responsibility, and comfort NNSs, it also raises important ethical concerns, particularly regarding the transparency of NNSs' communicative struggles. The automated self-disclosure of speaking difficulties, even when well-intentioned, involves revealing a form of vulnerability that NNSs may not have explicitly chosen to share. This concern may have already emerged among NNSs with high self-esteem and self-efficacy, who are hence reluctant to use speaking assistant tools.

Overall, the effectiveness of the mutual understanding channel underscores the importance of designing AI-mediated systems that not only support the linguistic performance of NNSs but also cultivate an inclusive communicative environment by actively engaging NSs. However, further research is warranted on the challenge of simultaneously achieving two goals: (1) reducing communicative barriers through enhanced mutual understanding, and (2) avoiding the exposure of language weaknesses that may cause discomfort for NNSs.

### 5.3 Design Implications

To address the multifaceted challenges proposed previously and promote the development of AI-mediated communication, we propose the following actionable design implications.

*5.3.1 Enhancing Support for Pronunciation and Conveyance of Translation Output.* While the speaking assistant aided NNSs in constructing expressions, some expressed difficulty in enunciating the translation outputs, potentially contributing to the lack of change in NNS's fluency and comfort (Section 4.4.1). The current speaking assistant cannot address this issue, suggesting a need for further refinements. Future research could introduce text-to-speech technology [62, 103] to directly verbalize or guide NNSs to pronounce the translated expression, as suggested by some NNSs in Section 4.3.1. Recent advances [103] in voice-generation technology now offer near-human quality, which may alleviate this discomfort, though further research is necessary to confirm its suitability in real-time communication. Another potential improvement is to enable automatic synchronization of translated text with both users' transcripts with visual indication (e.g., highlighted text), eliminating the need for NNSs to read it aloud. This approach could reduce reading pressure on NNSs while ensuring that NSs accurately capture key information.

*5.3.2 Alternative Methods for Exposing Linguistic Vulnerabilities.* A notable concern voiced by some NNS participants was a reluctance to use the speaking assistant in formal or evaluative settings, such as job interviews, academic presentations or conversations with superiors. This hesitation likely stems from a fear of exposing their linguistic limitations, which could be perceived negatively in such contexts. This indicates that while self-disclosure of language difficulties can effectively stimulate prosocial and supportive behavior from NSs in collaborative, lower-stakes environments, system designs must also consider the higher self-esteem [65, 82] and face-preserving needs of some NNSs, as overlooking these factors may inadvertently inhibit help-seeking behaviors. To address this concern, we suggest that future systems should explore alternative interaction paradigms. For instance, agent-mediated prompting could supplant explicit NNS help-seeking actions [74]. Instead of the NNS overtly activating the assistant, an AI agent could proactively detect linguistic hesitations, disfluencies [63, 96], or potential conceptual gaps in the NNS's speech. The agent could then discreetly suggest simplified rephrasings or clarifications to the NNS, while also subtly prompting NSs to slow their speaking pace or offer encouragement, without explicitly revealing the NNS's struggle or reliance on assistance.

*5.3.3 Facilitating NS-Initiated Clarification to Promote NNS Expression Refinement.* Compared to lower-proficiency NNSs, higher L2-proficiency individuals often avoid using the assistant, which may stem from confidence in their speaking clarity and efficiency [36]. However, this does not mean that NSs can truly understand these self-evaluated high proficiency NNSs, which risks a potential misunderstanding accumulation since NSs do not often actively seek NNSs to clarify the communication issues [31, 71, 87], even if this will greatly affect the smooth progress of the conversation. For example, Lindemann [71] found NSs with negative attitudes are more likely to use "avoidance strategies" towards NNSs, such as not providing feedback, not asking questions, and not seeking explanations, even if this may lead to task failure. Moreover, politeness norms or face-saving concerns [38, 44] are also critical considerations, suggested by the communication accommodation theory. This dynamic prevents mutual understanding and limits opportunities for NNSs' expression refinement. To address this, we recommend designing mechanisms that facilitate low-friction clarification requests from NSs. Specifically, NSs could be allowed to directly manipulate, such as select utterances from the transcript, and trigger a set of contextualized prompt buttons, such as *"Could you explain this?"*, *"Can you rephrase?"*, or *"What do you mean here?"*. Upon selection, the system would generate a light-touch, socially appropriate clarification prompt via LLM, encouraging the NNS to refine or re-express the idea. This approach allows NSs to express confusion without social discomfort, while also nudging NNSs, including high-proficiency ones, toward more precise communication and using speaking assistant in a positive way.

*5.3.4 Extending Conversational Support into Long-Term Learning Opportunities.* According to the language acquisition principles [14, 64, 101], challenges faced while participating in spoken conversation represent key learning opportunities to improve language skills. Currently, both high-proficiency NNSs and low-proficiency NNSs appear to be under-utilizing these learning opportunities. High-proficiency NNSs care about autonomy and face, thus missing the chance to achieve better expression through tools; lower-proficiency NNSs may be more troubled by a high cognitive load and achieving communication goals, making it difficult to identify suitable language development opportunities. Therefore, we recommend integrating a post-conversation module that automatically identifies moments where the speaking assistant was invoked or linguistic difficulties were detected. The system can then generate a review set with contextualized translations, vocabulary, grammar tips, and practice prompts for future rehearsal. By anchoring these learning materials to previously experienced communication scenarios, the review content becomes deeply embedded in meaningful conversational contexts [88] and linked to users' episodic memory [73], thereby enhancing retention and learning efficiency. This could be integrated with a spaced repetition engine to support cumulative learning over time.

## 6 Limitation and Future Work

This study has several limitations. First, this study was conducted in a controlled laboratory environment using a specific 'Objects Game' task to evaluate the impact of our system on NNS-NS communication. While this design enabled controlled comparison, such settings may not fully capture the complexity, spontaneity, and social dynamics of real-world communication, limiting ecological validity. Future studies could examine the system in more naturalistic tasks like open-ended conversations or workplace interactions to assess whether comparable benefits in self-efficacy and mutual understanding persist.

Second, the participant pool was limited to native Chinese speakers as NNSs and native English speakers as NSs. Although the demographics of the local community shaped this composition and provided meaningful insights into the speaking assistant's usability, the findings may not generalize across other language pairs or cultural contexts, where linguistic distance, pragmatics, and norms of interaction differ. Future work could extend this investigation to diverse linguistic pairings to better understand how such tools function across varied communicative contexts.

Third, our study focused on short-term interactions, capturing immediate improvements in self-efficacy and reduced anxiety. However, the potential long-term effects on language learning, communication patterns, and dependency on the speaking assistant remain unexplored. Longitudinal studies could investigate how sustained use shapes multilingual communication practices over time.

## 7 Conclusion

To help NNSs' speaking, we focus on mitigating their linguistic and interactional anxiety. Therefore, we propose a system that integrates voice-driven translation and a mutual understanding channel. Through a within-subjects experiments with 25-pair NNS-NS groups, our findings demonstrate its effectiveness in supporting NNSs during multilingual communication, manifested by the NNSs' enhanced speaking self-efficacy, decreased workload, and alleviated anxiety. Further analysis reveals that NNSs with lower L2-speaking proficiency levels can benefit more from the speaking assistant. Additionally, we show the essentiality of building mutual understanding, where NNSs' self-disclosed vulnerability raises the NSs' sense of sharing communication responsibility and empathy, and NSs' responses in turn help NNSs to engage with ease and lower anxiety about communication. Future research can explore broader applications and the long-term impact of such supportive tools on communication dynamics, as well as consider expanding the tool's adaptability for various linguistic and cultural backgrounds.

## References

[1] Norah Almusharraf and Daniel R Bailey. 2023. Students know best: Modeling the influence of self-reported proficiency, TOEIC scores, gender, and study experience on foreign language anxiety. *Sage Open* 13, 3 (2023), 21582440231179929.

[2] Seth Amoah and Joyce Yeboah. 2021. The speaking difficulties of Chinese EFL learners and their motivation towards speaking the English language. *Journal of Language and Linguistic Studies* 17, 1 (2021), 56–69.

[3] Michael E Anderson. 2014. Communication strategies and grounding in NNS-NNS and NS-NS interactions. (2014).

[4] Marta Antón and Frederick DiCamilla. 1998. Socio-cognitive functions of L1 collaborative interaction in the L2 classroom. *Canadian modern language review* 54, 3 (1998), 314–342.

[5] Ariyanti Ariyanti. 2016. Psychological factors affecting EFL students' speaking performance. *ASIAN TEFL Journal of Language Teaching and Applied Linguistics* 1, 1 (2016).

[6] Nooshan Ashtari. 2014. Non-native speech and feedback: The relationship between non-native speakers' production and native speakers' reaction. *The International Journal of Foreign Language Teaching* 9, 2 (2014), 9–17.

[7] J Baker. 2003. Essential Speaking Skills/Joanna Baker. *Westrup Heather: Continuum International Publishing Group* (2003).

[8] Albert Bandura. 2009. Social cognitive theory of mass communication. In *Media effects*. Routledge, 110–140.

[9] Albert Bandura and Sebastian Wessels. 1994. Self-efficacy. (1994).

[10] Muzakki Bashori, Roeland Van Hout, Helmer Strik, and Catia Cucchiarini. 2021. Effects of ASR-based websites on EFL learners' vocabulary, speaking anxiety, and language enjoyment. *System* 99 (2021), 102496.

[11] Elias Bensalem. 2018. Foreign Language Anxiety of EFL Students: Examining the Effect of Self-Efficacy, Self-Perceived Proficiency and Sociobiographical Variables. *Arab World English Journal* 9, 2 (2018).

[12] Joara Martin Bergsleithner. 2002. *Grammar and interaction in the EFL classroom: A sociocultural study*. Ph. D. Dissertation. Universidade Federal de Santa Catarina, Centro de Comunicação e Expressão ....

[13] Robert Berman and Liying Cheng. 2001. English academic language skills: Perceived difficulties by undergraduate and graduate students, and their academic achievement. *Canadian journal of applied linguistics* 4, 1 (2001), 25–40.

[14] Elsa Billings and Aida Walqui. 2018. The zone of proximal development: An affirmative perspective in teaching ELLs/MLLs. *Retrieved August* 6 (2018).

[15] Amber Bloomfield, Sarah C Wayland, Elizabeth Rhoades, Allison Blodgett, Jared Linck, Steven Ross, et al. 2010. What makes listening difficult? Factors affecting second language listening comprehension. *University of Maryland Center for Advanced Study of Language* (2010), 3–79.

[16] Virginia Braun and Victoria Clarke. 2006. Using thematic analysis in psychology. *Qualitative research in psychology* 3, 2 (2006), 77–101.

[17] J Brooke. 1996. SUS: A quick and dirty usability scale. *Usability Evaluation in Industry* (1996).

[18] Ruth Butler and Orna Neuman. 1995. Effects of task and ego achievement goals on help-seeking behaviors and attitudes. *Journal of educational Psychology* 87, 2 (1995), 261.

[19] Sabrina Chairunnisa and A Simangunsong Benedictus. 2017. Analysis of emoji and emoticon usage in interpersonal communication of Blackberry messenger and WhatsApp application user. *International Journal of Social Sciences and Management* 4, 2 (2017), 120–126.

[20] Mei-Ling Chen, Naomi Yamashita, and Hao-Chuan Wang. 2018. Beyond lingua franca: System-facilitated language switching diversifies participation in multiparty multilingual communication. *Proceedings of the ACM on Human-Computer Interaction* 2, CSCW (2018), 1–22.

[21] Charles Chiang and Diego Gomez-Zara. 2024. The Evolution of Emojis for Sharing Emotions: A Systematic Review of the HCI Literature. *arXiv preprint arXiv:2409.17322* (2024).

[22] Harald Clahsen and Claudia Felser. 2006. How native-like is non-native language processing? *Trends in cognitive sciences* 10, 12 (2006), 564–570.

[23] Herbert H Clark and Edward F Schaefer. 1987. Collaborating on contributions to conversations. *Language and cognitive processes* 2, 1 (1987), 19–41.

[24] Lacey Colligan, Henry WW Potts, Chelsea T Finn, and Robert A Sinkin. 2015. Cognitive workload changes for nurses transitioning from a legacy system with paper documentation to a commercial electronic health record. *International journal of medical informatics* 84, 7 (2015), 469–476.
[25] Lucy Coppinger and Sarah Sheridan. 2022. Accent Anxiety: An Exploration of Non-Native Accent as a Source of Speaking Anxiety among English as a Foreign Language (EFL) Students. *Journal for the Psychology of Language Learning* 4, 2 (2022), e429322.
[26] Henriette Cramer, Paloma De Juan, and Joel Tetreault. 2016. Sender-intended functions of emojis in US messaging. In *Proceedings of the 18th international conference on human-computer interaction with mobile devices and services.* 504–509.
[27] Jim Cummins. 2007. Rethinking monolingual instructional strategies in multilingual classrooms. *Canadian journal of applied linguistics* 10, 2 (2007), 221–240.
[28] Fred D Davis. 1989. Perceived usefulness, perceived ease of use, and user acceptance of information technology. *MIS quarterly* (1989), 319–340.
[29] Edward L Deci and Richard M Ryan. 1987. The support of autonomy and the control of behavior. *Journal of personality and social psychology* 53, 6 (1987), 1024.
[30] Robert M DeKeyser. 2005. What makes learning second-language grammar difficult? A review of issues. *Language learning* 55 (2005).
[31] Tracey M Derwing, Helen Fraser, Okim Kang, and Ron I Thomson. 2014. L2 accent and ethics: Issues that merit attention. *Englishes in multilingual contexts: Language variation and education* (2014), 63–80.
[32] Mark Dingemanse and Andreas Liesenfeld. 2022. From text to talk: Harnessing conversational corpora for humane and diversity-aware language technology. In *Proceedings of the 60th Annual Meeting of the Association for Computational Linguistics (Volume 1: Long Papers).* 5614–5633.
[33] Zoltán Dörnyei. 2014. *The psychology of the language learner: Individual differences in second language acquisition.* Routledge.
[34] Zoltán Dörnyei and Mary Lee Scott. 1997. Communication strategies in a second language: Definitions and taxonomies. *Language learning* 47, 1 (1997), 173–210.
[35] Wen Duan, Naomi Yamashita, and Susan R Fussell. 2019. Increasing native speakers' awareness of the need to slow down in multilingual conversations using a real-time speech speedometer. *Proceedings of the ACM on Human-Computer Interaction* 3, CSCW (2019), 1–25.
[36] David Dunning, Dale W Griffin, James D Milojkovic, and Lee Ross. 1990. The overconfidence effect in social prediction. *Journal of personality and social psychology* 58, 4 (1990), 568.
[37] Patricia M Duronto, Tsukasa Nishida, and Shin-ichi Nakayama. 2005. Uncertainty, anxiety, and avoidance in communication with strangers. *International Journal of Intercultural Relations* 29, 5 (2005), 549–560.
[38] Ali Elhami. 2020. Communication accommodation theory: A brief review of the literature. *Journal of Advances in Education and Philosophy* 4, 05 (2020), 192–200.
[39] Alan J Feely and Anne-Wil Harzing. 2003. Language management in multinational companies. *Cross Cultural Management: an international journal* 10, 2 (2003), 37–52.
[40] Claudia Felser. 2019. Structure-sensitive constraints in non-native sentence processing. *Journal of the European Second Language Association* 3, 1 (2019).
[41] Ana Gallego, Louise McHugh, Markku Penttonen, and Raimo Lappalainen. 2022. Measuring public speaking anxiety: self-report, behavioral, and physiological. *Behavior Modification* 46, 4 (2022), 782–798.
[42] Zhengdong Gan, Zi Yan, and Zhujun An. 2022. Development and validation of an EFL speaking self-efficacy scale in the self-regulated learning context. *Journal of Asia TEFL* 19, 1 (2022), 35.
[43] Ge Gao, Hao-Chuan Wang, Dan Cosley, and Susan R Fussell. 2013. Same translation but different experience: The effects of highlighting on machine-translated conversations. In *Proceedings of the sigchi conference on human factors in computing systems.* 449–458.
[44] Howard Giles. 1973. Accent mobility: A model and some data. *Anthropological linguistics* (1973), 87–105.
[45] Rebecca Godard and Susan Holtzman. 2022. The multidimensional lexicon of emojis: A new tool to assess the emotional content of emojis. *Frontiers in Psychology* 13 (2022), 921388.
[46] Agustín Gravano and Julia Hirschberg. 2011. Turn-taking cues in task-oriented dialogue. *Computer Speech & Language* 25, 3 (2011), 601–634.
[47] Dan W Grupe and Jack B Nitschke. 2013. Uncertainty and anticipation in anxiety: an integrated neurobiological and psychological perspective. *Nature Reviews Neuroscience* 14, 7 (2013), 488–501.
[48] William B Gudykunst. 2005. *Theorizing about intercultural communication.* Sage.
[49] Youssouf Haidara. 2016. Psychological Factor Affecting English Speaking Performance for the English Learners in Indonesia. *Universal Journal of Educational Research* 4, 7 (2016), 1501–1505.

[50] Ari Hautasaari and Naomi Yamashita. 2014. Catching up in audio conferences: highlighting keywords in ASR transcripts for non-native speakers. In *Proceedings of the 5th ACM international conference on Collaboration across boundaries: culture, distance & technology*. 107–110.

[51] Ari Hautasaari and Naomi Yamashita. 2014. Do automated transcripts help non-native speakers catch up on missed conversation in audio conferences?. In *Proceedings of the 5th ACM international conference on Collaboration across boundaries: culture, distance & technology*. 65–72.

[52] Helen Ai He, Naomi Yamashita, Ari Hautasaari, Xun Cao, and Elaine M Huang. 2017. Why did they do that? Exploring attribution mismatches between native and non-native speakers using videoconferencing. In *Proceedings of the 2017 ACM Conference on Computer Supported Cooperative Work and Social Computing*. 297–309.

[53] Pamela J Hinds, Tsedal B Neeley, and Catherine Durnell Cramton. 2014. Language as a lightning rod: Power contests, emotion regulation, and subgroup dynamics in global teams. *Journal of International Business Studies* 45 (2014), 536–561.

[54] Thomas Holtgraves and Caleb Robinson. 2020. Emoji can facilitate recognition of conveyed indirect meaning. *PloS one* 15, 4 (2020), e0232361.

[55] Elaine K Horwitz, Michael B Horwitz, and Joann Cope. 1986. Foreign language classroom anxiety. *The Modern language journal* 70, 2 (1986), 125–132.

[56] Peter Howell and Stevie Sackin. 2001. Function word repetitions emerge when speakers are operantly conditioned to reduce frequency of silent pauses. *Journal of psycholinguistic research* 30 (2001), 457–474.

[57] Xin Huang. 2023. A Review of Research on the Role of Translation in Second Language Acquisition. *Lecture Notes in Education Psychology and Public Media* 27 (2023), 201–205.

[58] Stuart A Karabenick. 2003. Seeking help in large college classes: A person-centered approach. *Contemporary educational psychology* 28, 1 (2003), 37–58.

[59] Stuart A Karabenick. 2004. Perceived achievement goal structure and college student help seeking. *Journal of educational psychology* 96, 3 (2004), 569.

[60] Ryan Kelly and Leon Watts. 2015. Characterising the inventive appropriation of emoji as relationally meaningful in mediated close personal relationships. In *Experiences of technology appropriation: Unanticipated users, usage, circumstances, and design*.

[61] Paul E King and Amber N Finn. 2017. A test of attention control theory in public speaking: Cognitive load influences the relationship between state anxiety and verbal production. *Communication Education* 66, 2 (2017), 168–182.

[62] Dennis H Klatt. 1987. Review of text-to-speech conversion for English. *The Journal of the Acoustical Society of America* 82, 3 (1987), 737–793.

[63] Tedd Kourkounakis, Amirhossein Hajavi, and Ali Etemad. 2021. Fluentnet: End-to-end detection of stuttered speech disfluencies with deep learning. *IEEE/ACM Transactions on Audio, Speech, and Language Processing* 29 (2021), 2986–2999.

[64] Stephen Krashen. 1982. Principles and practice in second language acquisition. (1982).

[65] Fiona Lee. 1997. When the going gets tough, do the tough ask for help? Help seeking and power motivation in organizations. *Organizational behavior and human decision processes* 72, 3 (1997), 336–363.

[66] Bawinda Sri Lestari, Joniarto Parung, and Frikson C Sinambela. 2021. Public speaking anxiety reviewed from self-efficacy and audience response on students: systematic review. In *International Conference on Psychological Studies (ICPSYCHE 2020)*. Atlantis Press, 75–81.

[67] Stephen C Levinson. 1983. *Pragmatics*. Cambridge university press.

[68] Jie Li, Ying Xia, Xinying Cheng, and Shijia Li. 2020. Fear of uncertainty makes you more anxious? Effect of intolerance of uncertainty on college students' social anxiety: A moderated mediation model. *Frontiers in Psychology* 11 (2020), 565107.

[69] Xiaoyan Li, Naomi Yamashita, Wen Duan, Yoshinari Shirai, and Susan R Fussell. 2023. Improving Non-Native Speakers' Participation with an Automatic Agent in Multilingual Groups. *Proceedings of the ACM on Human-Computer Interaction* 7, GROUP (2023), 1–28.

[70] John Lim and Yin Ping Yang. 2008. Exploring computer-based multilingual negotiation support for English–Chinese dyads: can we negotiate in our native languages? *Behaviour & Information Technology* 27, 2 (2008), 139–151.

[71] Stephanie Lindemann. 2002. Listening with an attitude: A model of native-speaker comprehension of non-native speakers in the United States. *Language in Society* 31, 3 (2002), 419–441.

[72] Leigh Anne Liu, Chei Hwee Chua, and Günter K Stahl. 2010. Quality of communication experience: Definition, measurement, and implications for intercultural negotiations. *Journal of Applied Psychology* 95, 3 (2010), 469.

[73] Lynn J Lohnas and M Karl Healey. 2021. The role of context in episodic memory: Behavior and neurophysiology. In *Psychology of Learning and Motivation*. Vol. 75. Elsevier, 157–199.

[74] Yaxi Lu, Shenzhi Yang, Cheng Qian, Guirong Chen, Qinyu Luo, Yesai Wu, Huadong Wang, Xin Cong, Zhong Zhang, Yankai Lin, et al. 2024. Proactive Agent: Shifting LLM Agents from Reactive Responses to Active Assistance. *arXiv*

*preprint arXiv:2410.12361* (2024).
[75] Kristina Lundholm Fors. 2015. Production and perception of pauses in speech. (2015).
[76] Peter D MacIntyre and Robert C Gardner. 1994. The subtle effects of language anxiety on cognitive processing in the second language. *Language learning* 44, 2 (1994), 283–305.
[77] Jiayue Mao. 2022. The Role of Nudges in Mitigating and Preventing Cyberbullying on Social Media. In *2022 3rd International Conference on Mental Health, Education and Human Development (MHEHD 2022)*. Atlantis Press, 1404–1408.
[78] Nunung Mardianti, Fitriani Rahayu, and Lalu Belik Made Dwipa. 2023. Between self-efficacy and speaking anxiety: A correlational study in non-english department students. *Scripta: English Department Journal* 10, 1 (2023), 175–181.
[79] Theresa Matzinger, Michael Pleyer, and Przemysław Żywiczyński. 2023. Pause Length and Differences in Cognitive State Attribution in Native and Non-Native Speakers. *Languages* 8, 1 (2023), 26.
[80] Nora McDonald, Sarita Schoenebeck, and Andrea Forte. 2019. Reliability and inter-rater reliability in qualitative research: Norms and guidelines for CSCW and HCI practice. *Proceedings of the ACM on human-computer interaction* 3, CSCW (2019), 1–23.
[81] Kristyan Spelman Miller. 2006. Pausing, productivity and the processing of topic in online writing. In *Computer Key-Stroke Logging and Writing*. Brill, 131–156.
[82] Arie Nadler. 1991. Help-seeking behavior: Psychological costs and instrumental benefits. https://api.semanticscholar.org/CorpusID:148967312
[83] Tsedal B Neeley. 2013. Language matters: Status loss and achieved status distinctions in global organizations. *Organization Science* 24, 2 (2013), 476–497.
[84] Helen M Newman. 1982. The sounds of silence in communicative encounters. *Communication Quarterly* 30, 2 (1982), 142–149.
[85] Tran Tin Nghi, Luu Quy Khuong, et al. 2021. A study on communication breakdowns between native and non-native speakers in English speaking classes. *Journal of English Language Teaching and Applied Linguistics* 3, 6 (2021), 01–06.
[86] Thi-Huyen Nguyen, Wu-Yuin Hwang, Xuan-Lam Pham, and Zhao-Heng Ma. 2018. User-oriented EFL speaking through application and exercise: Instant speech translation and shadowing in authentic context. *Journal of Educational Technology & Society* 21, 4 (2018), 129–142.
[87] Kayla Nymeyer, Dan P Dewey, William Eggington, and Wendy Baker-Smemoe. 2022. Factors that affect native English speakers' comfort levels when communicating with non-native English speakers. *International journal of applied linguistics* 32, 1 (2022), 158–174.
[88] Elinor Ochs and Bambi Schieffelin. 2001. Language acquisition and socialization: Three developmental stories and their implications. *Linguistic anthropology: A reader* 2001 (2001), 263–301.
[89] Hatice Okyar. 2024. Foreign language speaking anxiety and its link to speaking self-efficacy, fear of negative evaluation, self-perceived proficiency and gender. (2024).
[90] Mei Hua Pan, Naomi Yamashita, and Hao Chuan Wang. 2017. Task Rebalancing: Improving Multilingual Communication with Native Speakers-Generated Highlights on Automated Transcripts. In *the 2017 ACM Conference*.
[91] Yingxin Pan, Danning Jiang, Michael Picheny, and Yong Qin. 2009. Effects of real-time transcription on non-native speaker's comprehension in computer-mediated communications. In *Proceedings of the SIGCHI Conference on Human Factors in Computing Systems*. 2353–2356.
[92] Harold Pashler. 1994. Dual-task interference in simple tasks: data and theory. *Psychological bulletin* 116, 2 (1994), 220.
[93] S Passi and M Vorvoreanu. 2022. Overreliance on AI: literature review. Microsoft Research.
[94] Peinuan Qin, Zicheng Zhu, Naomi Yamashita, Yitian Yang, Keita Suga, and Yi-Chieh Lee. 2025. AI-Based Speaking Assistant: Supporting Non-Native Speakers' Speaking in Real-Time Multilingual Communication. arXiv:2505.01678 [cs.HC] https://arxiv.org/abs/2505.01678
[95] Pamela Rogerson-Revell. 2008. Participation and performance in international business meetings. *English for Specific Purposes* 27, 3 (2008), 338–360.
[96] Amrit Romana, Kazuhito Koishida, and Emily Mower Provost. 2024. Automatic disfluency detection from untranscribed speech. *IEEE/ACM Transactions on Audio, Speech, and Language Processing* (2024).
[97] Sherry Ruan, Jacob O Wobbrock, Kenny Liou, Andrew Ng, and James A Landay. 2018. Comparing speech and keyboard text entry for short messages in two languages on touchscreen phones. *Proceedings of the ACM on Interactive, Mobile, Wearable and Ubiquitous Technologies* 1, 4 (2018), 1–23.
[98] Emanuel A Schegloff. 1982. Discourse as an interactional achievement: Some uses of 'uh huh'and other things that come between sentences. *Analyzing discourse: Text and talk* 71 (1982), 71–93.
[99] Elizabeth Shriberg, Andreas Stolcke, and Don Baron. 2001. Observations on overlap: findings and implications for automatic processing of multi-party conversation.. In *Interspeech*. Citeseer, 1359–1362.
[100] XiaoLei Song, Yuan Cheng, and Ting Zeng. 2024. Using Cross-Cultural Machine Translation Technology to Promote Communication and Cooperation in English Courses. *Journal of Electrical Systems* 20, 6s (2024), 412–418.

[101] Merrill Swain. 1993. The output hypothesis: Just speaking and writing aren't enough. *Canadian modern language review* 50, 1 (1993), 158–164.

[102] Merrill Swain, Thomas Andrew KIRKPATRICK, and Jim Cummins. 2011. How to have a guilt-free life using Cantonese in the English class: A handbook for the English language teacher in Hong Kong. (2011).

[103] Xu Tan, Jiawei Chen, Haohe Liu, Jian Cong, Chen Zhang, Yanqing Liu, Xi Wang, Yichong Leng, Yuanhao Yi, Lei He, et al. 2024. Naturalspeech: End-to-end text-to-speech synthesis with human-level quality. *IEEE Transactions on Pattern Analysis and Machine Intelligence* (2024).

[104] Steven Taylor and Jaye Wald. 2003. Expectations and attributions in social anxiety disorder: Diagnostic distinctions and relationship to general anxiety and depression. *Cognitive Behaviour Therapy* 32, 4 (2003), 166–178.

[105] Helene Tenzer, Markus Pudelko, and Anne-Wil Harzing. 2014. The impact of language barriers on trust formation in multinational teams. *Journal of International Business Studies* 45 (2014), 508–535.

[106] Takahiro Tsumura and Seiji Yamada. 2023. Influence of agent's self-disclosure on human empathy. *PLoS One* 18, 5 (2023), e0283955.

[107] Saida Verdiyeva and Farida Huseynova. 2017. Speaking is a Challenging Skill in Language Learning. *International Journal of English Literature and Social Sciences* 2, 6 (2017), 239256.

[108] Felicidad T Villavicencio. 2011. Influence of self-efficacy and help-seeking on task value and academic achievement. *Philippine Journal of psychology* 44, 2 (2011), 166–180.

[109] Phuong Quyen Vo, Thi My Nga Pham, and Thao Nguyen Ho. 2018. Challenges to speaking skills encountered by English-majored students: A story of one Vietnamese university in the Mekong Delta. *CTU Journal of Innovation and Sustainable Development* 54, 5 (2018), 38–44.

[110] Sarah Wiseman and Sandy JJ Gould. 2018. Repurposing emoji for personalised communication: Why means "I love you". In *Proceedings of the 2018 CHI conference on human factors in computing systems*. 1–10.

[111] Cong Xu. 2025. The Effects of Anxiety on Self-Perceived Proficiency and Actual Achievement. *Open Journal of Social Sciences* 13, 3 (2025), 81–93.

[112] Naomi Yamashita, Andy Echenique, Toru Ishida, and Ari Hautasaari. 2013. Lost in transmittance: how transmission lag enhances and deteriorates multilingual collaboration. In *Proceedings of the 2013 conference on Computer supported cooperative work*. 923–934.

[113] Jing Ye and Tomoo Inoue. 2016. A speech speed awareness system for non-native speakers. In *Proceedings of the 19th ACM Conference on Computer Supported Cooperative Work and Social Computing Companion*. 49–52.

[114] Yuze Zeng, Junze Xiao, Danfeng Li, Jiaxiu Sun, Qingqi Zhang, Ai Ma, Ke Qi, Bin Zuo, and Xiaoqian Liu. 2023. The influence of victim self-disclosure on bystander intervention in cyberbullying. *Behavioral Sciences* 13, 10 (2023), 829.

# A Measurements

## A.1 Manipulation Check

NNS Questions:

(1) "Was the speaking assistant available in this room?"
(2) "Did the NS send an emoji/text response when you were using the tool (if the tool was unavailable, select No)?"

NS Questions:

(1) "Were you notified that the NNS was using a translation tool in this room?"
(2) "Did a response panel for emoji/text pop up when the NNS used the tool in this room?"

## A.2 NNS - Speaking Assistant Usability

(1) "I would like to use this speaking assistant frequently when speaking with NSs"
(2) "When I encountered an obstacle in speaking, I immediately thought of using the speaking assistant to help me overcome it"
(3) "I thought the speaking assistant was well integrated with the entire system and easy to use"
(4) "I would imagine that most people would learn to use this system very quickly".

## A.3 NNS - Speaking Assistant Usefulness

(1) "Using the speaking assistant gave me greater control over my speaking",
(2) "It would be difficult to communicate without the speaking assistant in some real-time multilingual communications with NSs",
(3) "Using the speaking assistant improved my speaking performance",
(4) "Using the speaking assistant saved me effort in speaking",
(5) "The speaking assistant made communication smoother and reduced awkward pauses",
(6) "I think it is unnecessary to introduce speaking assistant in the conversation (R)[9]",
(7) "Using the speaking assistant made my expressions more precise", and
(8) "Overall, I think the speaking assistant is very useful for the communication between the NS and me".

## A.4 NNS - Perception of the NS's Response

I thought that the NS's feedback...

(1) "alleviated my anxiety about speaking";
(2) "signaled to me that they were willing to wait for me to use the speaking assistant";
(3) "gave me a very positive feeling overall";
(4) "let me directly understand the NS's feelings, which eliminated some of my doubts and negative emotions".

## A.5 NNS - Speaking Self-Efficacy

(1) "I feel that I participated well in the process of completing the task with the NS"
(2) "I feel that I could keep up with the conversation and contribute meaningfully in these tasks"
(3) "I am confident that I could effectively communicate my ideas during the task".

(1) "I am confident in using the appropriate vocabulary to express my thoughts clearly in these tasks"

[9]R represents that the item is reverse-scored in the calculation of the scale.

(2) "I believe I can avoid misunderstandings by choosing the correct expressions and phrases"
(3) "I am confident in adjusting my expression to suit different conversational contexts during the task".

### A.6 NNS - Interactional Anxiety

(1) "I was worried that my speaking would disrupt the flow of the conversation"
(2) "I felt anxious that the NS will have to wait too long while I organize my thoughts"
(3) "I felt uneasy when I spoke slowly or paused during my expression".

### A.7 NNS - Workload

(1) **Mental demand**: "How much mental and perceptual activity was required in these tasks (e.g., thinking, deciding, calculating, remembering, looking, searching, etc.)?"
(2) **Temporal demand**: "How much time pressure did you feel due to the rate or pace at which the tasks or task elements occurred?"
(3) **Effort**: "How hard did you have to work (mentally and physically) to accomplish your level of performance?"
(4) **Frustration**: "How discouraged, irritated, stressed or annoyed did you feel?"

### A.8 NS - Evaluation of NNSs' Speaking Performance

Clarity:

(1) "I think the expression of NNS is very clear, and I can completely understand what they want to express"
(2) "I thought I didn't misunderstand what they want to express".

Comfort:

(1) "I feel very comfortable talking to my NNS partner, without any awkwardness caused by language"
(2) "I look forward to collaborating with my NNS partner on more tasks like this".

Fluency:

(1) "My NNS partner can express themselves fluently"
(2) "My NNS partner had very obvious pauses or hesitations when expressing themselves".

### A.9 NS - Motivation for Giving Response

(1) "I was sure that they encountered language difficulties at that time, so I wanted to offer some of my own support";
(2) "I believe that communication is not the sole responsibility of NNS; I still have a responsibility to contribute in my own way";
(3) "They looked nervous, and I thought that giving them some feedback may help alleviate their anxiety";
(4) "I can feel they are in a hurry and I want to give them some feedback to calm them down".

### A.10 Interviews

NNS Questions:

(1) "How did you use the speaking assistant? Did you use manual input or voice input, and why?"
(2) "Can you describe your experience using the speaking assistant? In which situations did you find it convenient and useful, and in which situations did you choose not to use it? What were your reasons?"

(3) "How did the NS's feedback impact your mental state while using the speaking assistant? Did you expect a response from the NS?"
(4) "Are there any specific areas in the current design that you feel need enhancement? Do you have any valuable suggestions for the future development of this tool?"

NS Questions:

(1) "How do you think the speaking assistant affected communication with the NNS?"
(2) "Do you find the time NNSs spent using the speaking assistant acceptable? What impact do you believe it had on your communication?"
(3) "Did you have any other reasons for responding to the notification that NNS was using the tool?"
(4) "How did you decide whether to use emojis or text in response to the notification that NNS is using the tool?"
(5) "Do you have any further thoughts on the current design?".

## B Tables

Table 1. Descriptive statistics of the speaking assistant's usability evaluation.

| Speaking Assistant Usability Description | Mean | SD |
|---|---|---|
| I would like to use this speaking assistant frequently when speaking with NSs | 5.000 | 1.588 |
| When I encountered an obstacle in speaking, I immediately thought of using the speaking assistant to help me overcome it | 5.583 | 1.213 |
| I thought the speaking assistant was well integrated with the entire system and easy to use | 5.625 | 1.173 |
| I would imagine that most people would learn to use this system very quickly | 6.125 | 0.680 |

Table 2. Descriptive statistics of the speaking assistant's usefulness evaluation.

| Speaking Assistant Usefulness Description | Mean | SD |
|---|---|---|
| Using the speaking assistant gave me greater control over my speaking | 5.458 | 0.884 |
| It would be difficult to communicate without the speaking assistant in some real-time multilingual communications with NSs | 4.958 | 1.732 |
| Using the speaking assistant improved my speaking performance | 5.417 | 1.100 |
| Using the speaking assistant saved me effort in speaking | 5.542 | 1.179 |
| The speaking assistant made communication smoother and reduced awkward pauses | 4.750 | 1.327 |
| I think it is unnecessary to introduce speaking assistant in the conversation (R) | 5.167 | 1.239 |
| Using the speaking assistant made my expressions more precise | 5.917 | 0.654 |
| Overall, I think the speaking assistant is very useful for the communication between the NS and me | 6.083 | 0.717 |

Table 3. Descriptive statistics of NNS perception of NS feedback.

| Perception Description | Mean | SD |
|---|---|---|
| I thought that NS's feedback alleviated my anxiety about speaking | 5.611 | 0.979 |
| I thought that NS's feedback signaled to me that they were willing to wait for me to use the speaking assistant | 5.722 | 1.018 |
| I thought that NS's feedback gave me a very positive feeling overall | 5.889 | 1.023 |
| I thought that NS's feedback let me directly understand the NS's feelings, which eliminated some of my doubts and negative emotions | 5.944 | 0.873 |

Table 4. Descriptive statistics of NS motivation to give feedback.

| Motivation Description | Mean | SD |
|---|---|---|
| I was sure that they encountered language difficulties at that time, so I wanted to offer some of my own support | 5.538 | 1.613 |
| I believe that communication is not the sole responsibility of NNS; I still have a responsibility to contribute in my own way | 5.846 | 1.345 |
| They looked nervous, and I thought that giving them some feedback may help alleviate their anxiety | 5.615 | 1.446 |
| I can feel they are in a hurry and I want to give them some feedback to calm them down | 5.308 | 1.494 |